\documentclass{aa}  

\usepackage{graphicx}
\usepackage{amsmath}
\usepackage{natbib}
\usepackage{txfonts}

\newcommand{\nside}{{\rm nside}}
\begin{document}

\title{Constraining the regular Galactic Magnetic Field with the 5-year WMAP
  polarization measurements at 22~GHz}

\titlerunning{GMF and WMAP5 data}
\subtitle{}

\author{ B. Ruiz-Granados\inst{1,2,3,4}, J.A.Rubi\~{n}o-Mart\'{i}n\inst{3,4}
  \and E. Battaner\inst{1,2} }

\authorrunning{Ruiz-Granados et al.}

\institute{
Dpto. F\'{i}sica Te\'{o}rica y del Cosmos. Edif. Mecenas, planta baja, 
Campus Fuentenueva, E-18071. Universidad de Granada, Granada(Spain)  
\and Instituto de F\'{i}sica Te\'{o}rica y Computacional Carlos I,
Granada(Spain)  
\and Instituto de Astrof\'{i}sica de Canarias (IAC), C/V\'{i}a L\'{a}ctea, s/n,
E-38200, La Laguna, Tenerife (Spain) 
\and Departamento de Astrof\'{i}sica, Universidad de La Laguna, E-38205, La Laguna, Tenerife (Spain)  }

\offprints{B. Ruiz-Granados, \email{bearg@iac.es}. }

\date{Received / Accepted}

\abstract {The knowledge of the regular (large scale) component of the Galactic
  magnetic field gives important information about the structure and dynamics of
  the Milky Way, as well as constitutes a basic tool to determine cosmic rays
  trajectories. It can also provide clear windows where primordial magnetic
  fields could be detected. }
{We want to obtain the regular (large scale) pattern of the magnetic field
  distribution of the Milky Way that better fits the polarized synchrotron
  emission as seen by the WMAP satellite in the 5 years data at 22~GHz.}
{We have done a systematic study of a number of Galactic magnetic field models:
  axisymmetric (with and without radial dependence on the field strength),
  bisymmetric (with and without radial dependence), logarithmic spiral arms,
  concentric circular rings with reversals and bi-toroidal. We have explored
  the parameter space defining each of these models using a grid-based
  approach. In total, more than one million models are computed.  The model
  selection is done using a Bayesian approach. For each model, the posterior
  distributions are obtained and marginalised over the unwanted parameters to
  obtain the marginal (one-parameter) probability distribution functions. }
{In general, axisymmetric models provide a better description of the
    halo component, although attending to their goodness-of-fit, the rest of the
    models cannot be rejected.  In the case of disk component, the analysis is
    not very sensitive for obtaining the disk large scale structure, because of
    the effective available area (less than 8\% of the whole map and less than
    40\% of the disk). Nevertheless, within a given family of models, the
    best-fit parameters are compatible with those found in the literature. }
{The family of models that better describes the polarized synchrotron
    halo emission is the axisymmetric one, with magnetic spiral arms with a
    pitch angle of $\approx 24^{\circ}$, and a strong vertical field of 1~$\mu$G
    at $z \approx 1$~kpc. When a radial variation is fitted, models require fast
    variations. }


\maketitle

%

%
\section{Introduction}
\label{sec01}

Spiral galaxies exhibit large-scale magnetic fields. The Milky Way is not an
exception, but obtaining its spatial distribution is extremely difficult.
Most methods for observing magnetic fields are based either on radio
observations of the synchrotron emission \citep[][and references
therein]{wolleben06,reich06,testori08}, or on the Faraday Rotation (hereinafter,
FR) of pulsars \citep[e.g.][]{weisberg04,han06,noutsos08} and extragalactic
radio sources (hereinafter, EGRS)
\citep[e.g.][]{gaensler01,brown07,haverkorn08,carretti08}.
In addition, several radio lines show Zeeman splitting, which can be used as
well to directly constrain the strength of the magnetic field
\citep[e.g.][]{fish03,han07}.

Despite the large numbers of pulsars and EGRS for which the Rotation Measure
(hereinafter, RM) has been recently determined, there is no consensus about the
large-scale pattern of the Galactic Magnetic Field (GMF).
Probably, it is more complex than previously expected, as pointed out in
\cite{men08}. Recent results by \cite{sun08} show an axisymmetric disk
distribution with reversals inside the solar circle as the best model to
describe the GMF when all-sky maps at 1.4~GHz from DRAO and Villa Elisa, the
22~GHz map from WMAP satellite, and the Effelsberg RM survey of EGRS are
combined. Results derived from~\cite{brown07} by using RM of EGRS suggest an
axisymmetric pattern of the disk magnetic field. \cite{vallee08} claimed for an
inclusion of a ring model to describe the field.
%
The interest on the structure of this large scale GMF is justified by a number
of reasons. First of all, this field might be of importance when considering the
dynamics of the galaxy at those large scales
\citep{nelson88,battaner92,battaner95,kutschera04,battaner07}.
A good characterization of the large-scale GMF pattern would allow a detailed
correction of the galactic contribution for a better understanding of
cosmological magnetic fields \citep[see ][for a recent review]{battaner09},
which would be potentially observed with upcoming CMB missions, such as PLANCK
\citep{planck06}.

In addition, the GMF modifies the trajectory of high energy cosmic rays,
therefore its knowledge is crucial to understand their distribution in energy
and direction, such as obtained in experiments as {\sc Auger} \citep{auger08},
{\sc Milagro} \citep{abdo09} and others.
The global anisotropies found by {\sc Milagro} have been interpreted by
  \cite{battaner09let} as produced by galactic magnetic fields.
Moreover, GMF could be important to explain the well known knee in the spectrum
of cosmic rays energies around $10^{6}$~GeV \citep{masip08}, as sub-knee-energy
cosmic protons trapped by GMF should have a much larger optical depth for
interactions with WIMP's.
%

The extraction of structure of the galactic magnetic field from the measurements
of the polarized intensity is extremely difficult. The galactic magnetic field
cannot probably be described by a single model but there could be at least three
components, the thin disk, the thick disk \citep{beuermann85} and the halo, each
one defined by its own model and parameters. Taking into account that each
component is largely unknown, an analysis of a multi-component magnetic field
renders an objective estimation of the best configuration a too complicated
task.

The thin disk is characterized by the highest field strengths and we are
embedded in it. However, magnetic fields in the thin disk are
  predominantly random, and therefore they barely contribute to the observed net
  polarized emission, even if a $z$-component of the regular fields exist
  \citep{han99}. On the other hand, local spurs \citep{berkhuijsen71}, as the
North Polar Spur, highly distort the main field configuration. Even at this high
frequency Faraday depolarization cannot be neglected in particular
regions. Observing at high galactic latitudes the thin disk field is
contaminated by the thick disk and the halo fields, which could even become
dominant. Models for the thick disk are scarce in the literature but those
proposed for the thin disk could be tested, even if different parameters could
characterize both disks. Our results for the disk inferred from the
  polarized synchrotron emission will give a complementary insight on the field
  structure. The halo structure remains unknown and has a very different
  structure consisting in a double torus in two hemispheres with opposite
  directions. After some pioneer detections \citep{simard80,han94}, it has
been modelled by \cite{han97},~\cite{harari99}, \cite{tinyakov02},
\cite{prouza03} and others. To illustrate the current uncertainty on this
component, we can compare the maximum value of the magnetic strength, 1~$\mu$G
following Prouza and Smida, and 10~$\mu$G following Sun et al., although these
last authors propose a maximum strength of only 2~$\mu$G when the thermal
electron scale height is increased by a factor of 2. The contribution of the
halo field to the polarized emission is therefore difficult to estimate. Most
models take only into account RM from extragalactic sources to estimate the halo
structure.


%
%
In this work, we carry out a systematic comparison of a number of possible GMF
models, exploring which one is providing a better fit to the large-scale
polarization map at 22~GHz.

Our analysis is based on the 5-year WMAP data release, and extends the
work by \cite{page07} by considering a detailed comparison not only with the
polarization angle, but with the polarized intensity (i.e. Stokes's Q and U
parameters).
Although the polarization angle can be used to described some properties of the
large scale pattern of the GMF, it is not sensitive to some parameters (e.g. the
field strength) and it also contains some intrinsic degeneracy with respect to
the direction of the field lines. Because of this reason, our main results will
be obtained with the analysis of the $(Q,U)$ maps, although the independent
analysis based on the position angle will be done in some cases for comparison.

%
%
In this work, we have used eight disk models and one halo
model. Different masks enable us, in an indirect way, to estimate the different
contributions of the galactic components at different galactic latitudes, but
the consideration of a multicomponent field has not been fully undertaken. In
what follows, we will speak of the disk without specifying if thin or thick.

For obtaining the emission we also need a model of the distribution and spectrum
of cosmic rays which is another important source of uncertainties. Here we have
assumed that the cosmic ray structure follows that of the gas, as they are
produced by supernovae and cannot travel faraway from the birth place. This
assumption is rather usual in the literature but rather questionable too.

Here we focus on the study of the large scale field. In principle, we could
divide the galactic magnetic field into three components: a random component for
scales lower than say 100~pc \citep[see][]{haverkorn08}, for which some
works have published the turbulence spectrum \citep{han04,han09}; a
main ``spiral'' field for scales typical of spiral arms, and a ``galactic
scale'' main field. Some authors consider that 1~kpc is a length defining the
large scale \citep[e.g.][]{han08}, therefore taking spiral arms as a large scale
phenomenon. In fact, the field direction is opposite in arms and in inter-arms
regions \citep[e.g.][]{beck96,han06}.
Here, however, we are interested in scales of the galaxy itself, being therefore spiral arms
considered as wavy perturbations.  It is clear that we need to investigate these
three components, but this separation is important because the tools, and mainly
the interpretation in terms of the generation mechanism, could be completely
different. The random field should be produced by turbulence in a magnetized
medium, supernova explosions and other local mechanisms. Spiral arms produce
characteristic motions that enhance magnetic fields in a high conductivity
medium. Fields at the galactic scale would be interpreted in terms of galaxy
formation and/or dynamo effects.

The polarized synchrotron emission is a valuable tool for investigating
  the overall field pattern at large scales and specially at high galactic
  latitudes. The total emission is much affected by random fields and by
  non-polarized galactic emission as free-free \citep[see
  e.g.][]{miville08}. FR of pulsars is a powerful technique but it is very much
affected by enhancements and tangling of the field by the passage of spiral
waves. FR of extragalactic sources would inform about the larger scale
fields but here the poor knowledge of the intrinsic FR is a heavy problem. When
using all-sky observations such as provided by WMAP, and PLANCK in the future,
the observed emission is integrated along a path traversing the galaxy,
therefore the wavy effect of spiral arms is, at least in part, smoothed.

This could explain why RM of pulsars and WMAP polarization have given somewhat
inconsistent results. There are other all-sky surveys at lower frequencies
\citep[see][]{reich06} but Faraday depolarization and dust emission are very
strong at latitudes below $30\deg$. The detailed study of the WMAP polarization
measurements in terms of the galactic scale magnetic fields is more a
complementary than an additional technique. It is relatively free of Faraday
depolarization, of dust contamination, of random fields and of spiral waves
perturbations.


To complement our study, for some of the models we also investigate if a
modification of the radial variation of the strength of the GMF is producing an
impact on the quality of the fit. This radial variation is also largely unknown
and may be of importance on the production of the all-sky polarization maps.


The paper is organized as follows. In Section~2 we describe the set of GMF
models that have been investigated in this analysis. In Section~3, we describe
the numerical method used to produce the template maps, as well as the the model
selection method used to determine our preferred model and the corresponding set
of parameters.  We present the results in Section~4 and discuss them in the
following section.

%
\section{Models of Galactic Magnetic Field}
\label{sec:02}

We have focused our analyses on the constraints derived from the polarized
intensity map at 22~GHz. At these frequencies, the physical process which
dominates the polarized intensity is the synchrotron radiation emitted by the
population of relativistic cosmic ray (CR) electrons with energies between
400~MeV and 25~GeV \citep{strong07}, as they interact with the GMF.

Therefore, in order to obtain a prediction of the polarization pattern at this
frequency, we first require a description of the distribution of the
relativistic CR electrons in our Galaxy. For the purposes of this paper, in
which we are interested only in the large-scale pattern of the GMF, it is
sufficient to consider a simplified description of the CR electron population.
Here, we use the spatial distribution of relativistic electrons as in
\cite{drimmel01}, i.e.
\begin{equation}
\label{ncre}
N_{e} = N_{0} \exp\left( \frac{-r}{5~{\rm kpc}} \right) sech^{2}\left(
\frac{z}{1~{\rm kpc}} \right)
\end{equation}
where $N_{0} \approx 3.2 \times 10^{-4} cm^{-3}$ is derived from the value for
the CR electron density on Earth \citep{sun08}, and $r$ and $z$ are the radial
and the vertical coordinates in cylindrical galactocentric coordinates,
respectively.
This model essentially assumes the same spatial density distribution for the CR
electrons as for the interstellar gas. At first order, this is what we would
expect, because the higher the interstellar gas density, the higher the star
formation rate, and therefore, the higher supernova production which gives a
higher relativistic electron density. These cosmic electrons lose energy (via
synchrotron itself) in short distances of less than 1~kpc.

The value for $N_0$ in equation~(\ref{ncre}) is very uncertain.  This value is
usually obtained by assuming that the CR electron spectrum can be described by a
power-law with constant spectral index $p=3$. However, observations in the last
few years, as well as numerical simulations \citep[see ][and references
therein]{strong07} suggest that this assumption is not appropriate for the
entire spectrum. Moreover, different observations show variations of the order
of 50\% (or even larger) for this quantity.
Therefore, we expect that this uncertainty might introduce a bias in the
recovered amplitude of the field strength for the different models, and will be
taken into account as explained below.

In the following sub-sections we present the set of GMF models we investigate in
this work, all of them taken from the literature. Most of these models have been
proposed/constrained using the analysis of Faraday Rotation of pulsars and
EGRSs. Thus, it is also interesting to explore if these models are also able to
describe the large-scale polarization pattern seen in WMAP 22~GHz maps.
In total, we investigate eight models describing the disk and
  halo fields, and one model conceive to describe the halo field. The
disk models are: 1) Axisymmetric (ASS), 2) Axisymmetric with radial dependence
of the strength (ASS(r)), 3) Bisymmetric positive (BSS$_{+}$) , 4)
  Bisymmetric negative (BSS$_{-}$), 5) Bisymmetric positive with radial
  dependence of the strength (BSS$_{+}$(r)), 6) Bisymmetric positive with radial
  dependence of the strength (BSS$_{+}$(r)), 7) Concentric Circular
Rings (CCR) and 8) Logarithmic Spiral Arms (LSA). Moreover,
for the halo component, we shall consider the bi-toroidal (BT) model. Throughout
this section, all coordinates ($r,z$) refer to cylindrical galactocentric
coordinates.

\subsection{Axisymmetric Model}

The axisymmetric model \citep[see e.g.][]{vallee91,poezd93} is one of the
simplest descriptions of the GMF. It is compatible with a non-primordial origin
of the galactic magnetism, based on the dynamo theory. There are several
possible models of the family of the axisymmetric distribution.  The components
for this model of GMF are given by:
\begin{subequations}
  \label{eq:ASS}
  \begin{align}
    \label{eq:ASS_r}
     B_{r} &= B_{0}(r) \sin(p) \cos(\chi(z))\\
    \label{eq:ASS_phi}
     B_{\phi} &= B_{0}(r) \cos(p) \cos(\chi(z)) \\
    \label{eq:ASS_z}
     B_{z} &= B_{0}(r) \sin(\chi(z))
  \end{align}
\end{subequations}
where $p$ is the pitch angle\footnote{The ``pitch angle'' is defined
    here as the angle between the azimuthal direction and the magnetic
    field. The azimuthal direction ($\hat{\phi}$) increases in anti-clockwise,
    so $p$ is positive since the anti-clockwise tangent to the spiral is outside
    the circle with radius r. Note that in some works in the literature, the
    convention for the $\hat{\phi}$-angle is exactly opposite (i.e. it increases
    clockwise). Finally, note that for the case $p = 0^{\circ}$, the solenoidal
    model is recovered.} which it is considered constant; $B_{0}(r)$ is the
field strength (which in principle might be a function of the radial distance),
and $\chi(z)$ corresponds to a ``tilt angle''. For our study, we have adopted
the following functional dependence:
\begin{equation}
 \chi(z)= \chi_{0}\tanh(\frac{z}{z_0})
\label{eq:tilt}
\end{equation}
where we use a value of $z_0=1~{\rm kpc}$ for the characteristic scale of
variation in the vertical direction.

For the computations in this paper, we shall consider two cases for the radial
dependence of $B_0(r)$. The first case (hereafter ASS) corresponds to $B_0(r) =
B_0$ (constant value).  As a second case (hereafter ASS(r)), we consider a radial
variation of the type:
\begin{equation}
B_{0}(r)=\frac{B_{1}}{1+\frac{r}{r_{1}}}
\label{eq:radial}
\end{equation}
where $r_1$ represents the characteristic scale at which $B_{0}(r)$ decreases to
half of its value at the galactic centre. This radial variation is based on a
possible extension of the model by \cite{poezd93}.
The expression has appropriate asymptotic behaviors, in the sense that we obtain
a finite value when $r$ is close to the galactic center ($r \rightarrow 0$), and
asymptotically tends to $\propto 1/r$ when $r \rightarrow \infty$, as suggested
in \cite{battaner07}.
We must note that in equation~(\ref{eq:radial}), $B_{1}$ and $r_{1}$ are not
independent, provided that we fix the value of the GMF strength in the solar
neighbourhood. For example, using $R_0=8$~kpc for the Sun galactocentric
distance, and $B_\odot= 3$~$\mu$G for the magnetic field strength in the solar
neighbourhood, we can re-write equation~(\ref{eq:radial}) in terms of a single
free-parameter as:
\begin{equation}
B_{0}(r) = \frac{3r_{1}+24}{r_{1}+r}
\label{eq:radial2}
\end{equation}
where $r$ is given in kpc and $B_{0}(r)$ in $\mu$G.

Summarizing, a particular ASS model is fully described once these three
parameters are given: $[ B_0, p, \chi_{0}]$, while the ASS(r) model, one would in
principle require four parameters. However, for the ASS(r) family of models, we
will use the constraint given in equation~(\ref{eq:radial2}), which in practice
means that we will only have three free parameters to represent a certain model:
$[r_{1}, p, \chi_{0}]$.

The typical range of variation of the $p$ values found in the literature for the
ASS model of the disk \citep[see ][]{vallee91} is shown in Table~1. In general, different
pitch angle and field strength values are derived for the different spiral arms
of the Galaxy.

\subsection{Bisymmetric Model}
This model is compatible with a primordial origin of the cosmic magnetism. It
could explain the reversals of the magnetic field derived from observations of
RM of pulsars \citep[e.g.][]{han94,han01,han06}. The three components for
this model are given by:
\begin{subequations}
  \label{eq:BSS}
  \begin{align}
    \label{eq:BSS_r}
     B_{r} &= B_{0}(r)\cos \left( \phi \pm \beta \ln \left( \frac{r}{R_{0}}
     \right) \right) \sin(p)\cos(\chi(z)) \\
    \label{eq:BSS_phi}
     B_{\phi} &= B_{0}(r)\cos \left( \phi \pm \beta \ln \left( \frac{r}{R_{0}}
     \right) \right) \cos(p) \cos(\chi(z))\\
    \label{eq:BSS_z}
     B_{z} &= B_{0}(r) \sin(\chi(z))
  \end{align}
\end{subequations}
where $B_{0}(r)$ is the field strength; $\beta = 1/\tan(p)$, being $p$ the pitch
angle; $R_{0}$ is the distance Sun-Galactic center ($\approx 8$~kpc); and
finally $\chi(z)$ is the tilt angle, which we assume is also given by
equation~(\ref{eq:tilt}). Note that in Eq.~(\ref{eq:BSS}) we are considering
  two possible families of bisymmetric models, which correspond to the positive
  (negative) sign inside the big parenthesis for $B_r$ and $B_\phi$. In this
  way, we are considering the two conventions that are found in the
  literature. ``Positive'' bisymmetric models (hereafter, BSS$_{+}$) are defined
  with the same convention to recover the model proposed by \cite{han94}, while
  ``negative'' bisymmetric models (hereafter, BSS$_{+}$) use the same sign
  convention of \cite{jansson09}. Note that for BSS$_{-}$, the spiral has
  opposite sign for the magnetic direction.

As in the previous case, we consider two sub-families of models. The first one,
noted as BSS$_{\pm}$, corresponds to constant strength of the
magnetic field. In this case, a model is completely defined by giving three
parameters: $[B_{0}, p, \chi_{0}]$.
The second family, noted as BSS$_{\pm}$(r), includes a radial variation
of the field strength according to equation~(\ref{eq:radial}). For this case, we
again fix the field strength to the value in the Solar neighbourhood
(equation~\ref{eq:radial2}), so a model is specified by giving three parameters:
$[r_{1}, p, \chi_{0}]$.

Typical parameter values for the disk model found in the literature, after converted to our convention for the sign of the pitch angle, are $p = (8.2 \pm{0.5})^{\circ}$, $B_{0} = 1.4$~$\mu$G
and $B_{global} = (1.8 \pm{0.3})$~$\mu$G \citep{han94}; $p=14^{\circ}$ and
$B_{0} = 1$~$\mu$G \citep{simard80}; and $p = 11^{\circ}$ and $B_{0} =
(2.1\pm{0.3})$~$\mu$G \citep{han06}.


\subsection{Concentric Circular Ring (CCR) model }

This model was proposed by \cite{rand89} to fit the RM of their pulsars
catalogue, with the aim of providing a way to take into account reversals of the
magnetic field at different radii. In their original expressions, they do not
consider a vertical component for the magnetic field (i.e. $B_z=0$). Here, we
shall extend the equations for this model presented in \cite{indrani99} to
account for a vertical dependence, in the following way:
\begin{subequations}
  \label{eq:CCR}
  \begin{align}
    \label{eq:CCR_r}
     B_{r} &= 0\\
    \label{eq:CCR_phi}
     B_{\phi} &= \frac{B_{0}}{ \sin(\pi D_r/w)} \sin \Big(  \frac{\pi (r - R_{0} + D_{r})}{w} \Big) \cos(\chi(z))\\
    \label{eq:CCR_z}
     B_{z} &= B_{0} \sin(\chi(z)) 
  \end{align}
\end{subequations}
where $w$ is the space between reversals; $D_{r}$ is the distance to the first
reversal; $B_{0}$ is the field strength, and finally $\chi(z)$ is given by
equation~(\ref{eq:tilt}).  All distances ($w$ and $D_r$) are given in kiloparsecs.
Note that we introduce an additional factor $\sin(\pi D_r/w)$ in the definition
of $B_\phi$, so $B_0$ still preserves the meaning of the magnetic field strength
in the solar neighbourhood.

In our analyses, we have not considered any radial dependence of the field
strength, so for the CCR model, the parameter space is defined by $[D_{r},w,
  B_{0}, \chi_{0}]$.

The best-fit values obtained by \cite{rand89} for the disk component were: $D_{r}=(0.6\pm{0.08})$~kpc ,
$w = (3.1\pm{0.08})$~kpc and $B_{0}= (1.6 \pm{0.2})$~$\mu$G.

\subsection{Logarithmic Spiral Arms (LSA) model}

This model was used by \cite{page07} to describe the distribution of the
large-scale pattern of the polarization angle in the 22~GHz WMAP 3-year data. The
equation which describes the field is:
\begin{subequations}
  \label{eq:Page}
  \begin{align}
    \label{eq:Page_r}
     B_{r} &= B_{0}\sin\psi(r)\!\cos\chi(z)\\
    \label{eq:Page_phi}
     B_{\phi} &= B_{0}\cos\psi(r)\!\cos\chi(z)\\
    \label{eq:Page_z}
     B_{z} &= B_{0}\sin\chi(z) 
  \end{align}
\end{subequations}
where
\[ \psi(r)=\psi_{0}+\psi_{1}ln \left( \frac{r}{8~{\rm kpc}} \right) \]
and
\[
\chi(z)=\chi_{0}tanh\left( \frac{z}{1~{\rm kpc}} \right)\]
Note that the LSA model is essentially an axisymmetric model in which the pitch
angle is not constant, being magnetic field lines logarithmic spirals.
According to our definition of pitch angle for the ASS model, $\psi(r)$ would
play the role of pitch angle, in which we would have a constant part,
$\psi_{0}$, and a characteristic amplitude for the logarithmic dependence of the
arms, $\psi_{1}$.
Following \cite{page07}, $B_{0}(r)$ is assumed constant, with a value of
$3$~$\mu$G.

Thus, the parameter space is defined by $[\psi_{0}, \psi_{1}, \chi_{0}]$. The
proposed values for the different parameters are $[\psi_{0}, \psi_{1}, \chi_{0}]
= [ 27^{\circ}, 0.9^{\circ}, 25^{\circ}]$ \citep[see][and also the erratum
  available at the LAMBDA web
  site\footnote{http://lambda.gsfc.nasa.gov/product/map/current/map\_bibliography.cfm}]{page07}.

\subsection{Bi-Toroidal (BT) model (halo model)}

Some authors have found hints of a halo or thick disc field component with
opposite directions in both hemispheres. For example, \cite{han02} and
\cite{prouza03} detected it with maximum strengths at a large height of
$1.5$~kpc at both sides above and below the plane, being the maximum strengths
of $1$~$\mu$G approximately. \cite{sun08} give a more complete description of
this double torus, being the height of maximum strength again at $1.5$~kpc, but
its maximum is much stronger (around $10$~$\mu$G).

Following this scenario, we propose a possible configuration for the components
of the magnetic field which do contain a different sign in both hemispheres, and
it is given by:
\begin{subequations}
  \label{eq:TD}
  \begin{align}
    \label{eq:TD_r}
     B_{r} &= 0\\
    \label{eq:TD_phi}
     B_{\phi} &= B_{0}(r) \arctan \left( \frac{z}{\sigma_{1}} \right) \exp \left( \frac{-z^{2}}{2 \sigma_{2}^{2}} \right)\\
    \label{eq:TD_z}
     B_{z} &= {\rm constant}
  \end{align}
\end{subequations}
where $\sigma_{1}$ and $\sigma_{2}$ are two constants (measured in kpc) which
encode the characteristic scales of variation of the field with the vertical
distance, and do take into account in a simplified way the change in the sign.
For our computations, we fix the $B_z$ value to be $0.2$~$\mu$G \citep{han94},
and we only consider the case of a radial variation of $B_0(r)$ given by
equation~(\ref{eq:radial2}). Thus, the parameter space for this model is specified
by $[r_{1}, \sigma_{1}, \sigma_{2}]$.

\section{Methodology}
In order to obtain constraints on the different parameters for each one of the
models described above, we have performed a systematic comparison of the
predicted polarized intensity due to synchrotron emission with the observed map
at 22~GHz by WMAP satellite.
Here, we describe the relevant details of the dataset, the numerical procedure
to compute the synchrotron maps for a certain GMF model, and the model selection
criterion that we have adopted for our analyses.

\subsection{Description of the K-band WMAP5 data}
\label{sec:wmapdata}

The analysis of this paper is based on a comparison with the K-band (equivalent
to 22~GHz) polarization map obtained by the WMAP satellite after five years of
operation \citep{hinshaw09}.  This map is publicly available in the LAMBDA
website\footnote{http://lambda.gsfc.nasa.gov/product/map/current}, and it is
given in {\sc HEALPix}\footnote{http://healpix.jpl.nasa.gov/} format
\citep{gorski05}.

\begin{figure}
\centering
\includegraphics[height=0.9\columnwidth,angle=90]{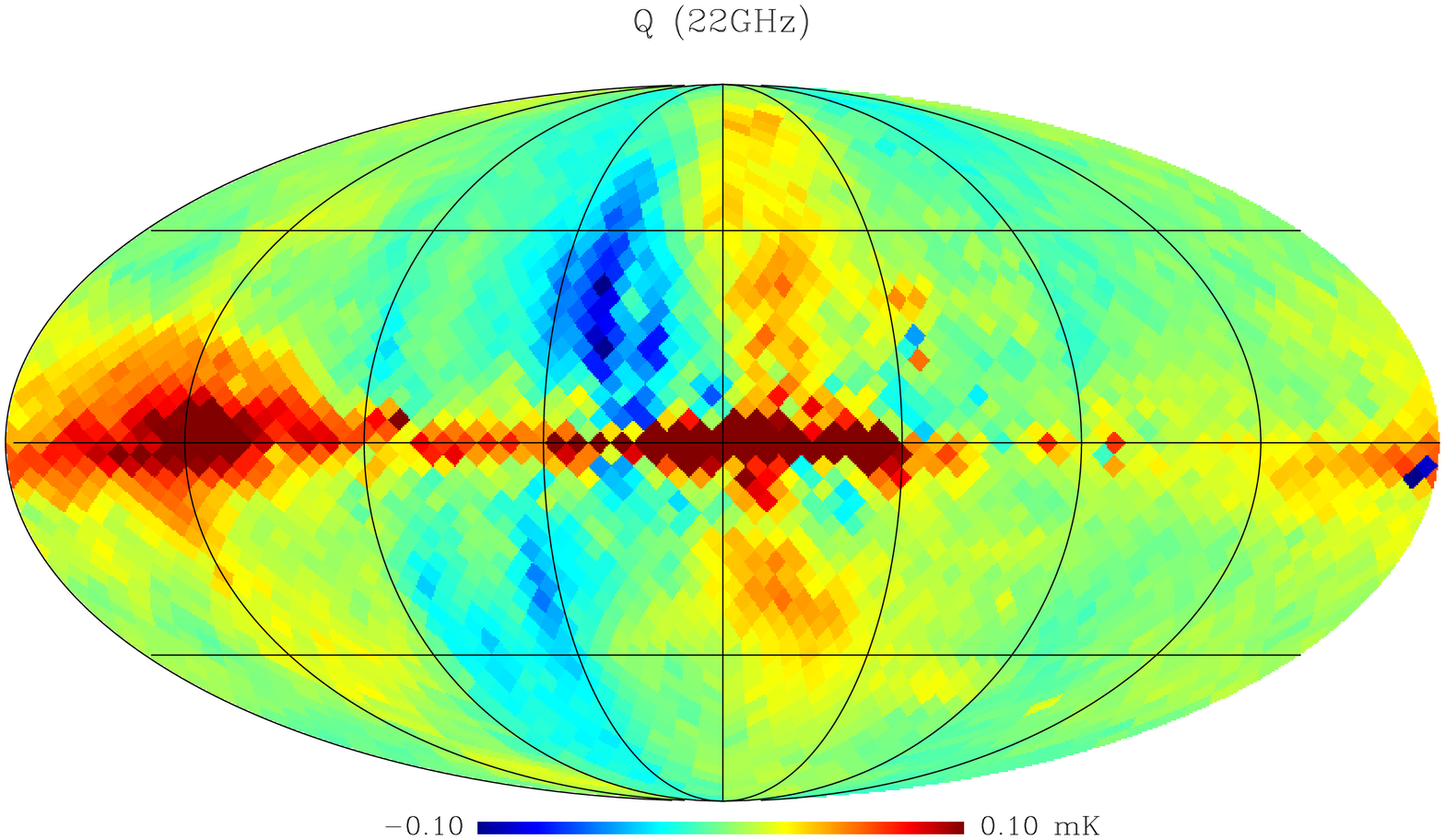}
\includegraphics[height=0.9\columnwidth,angle=90]{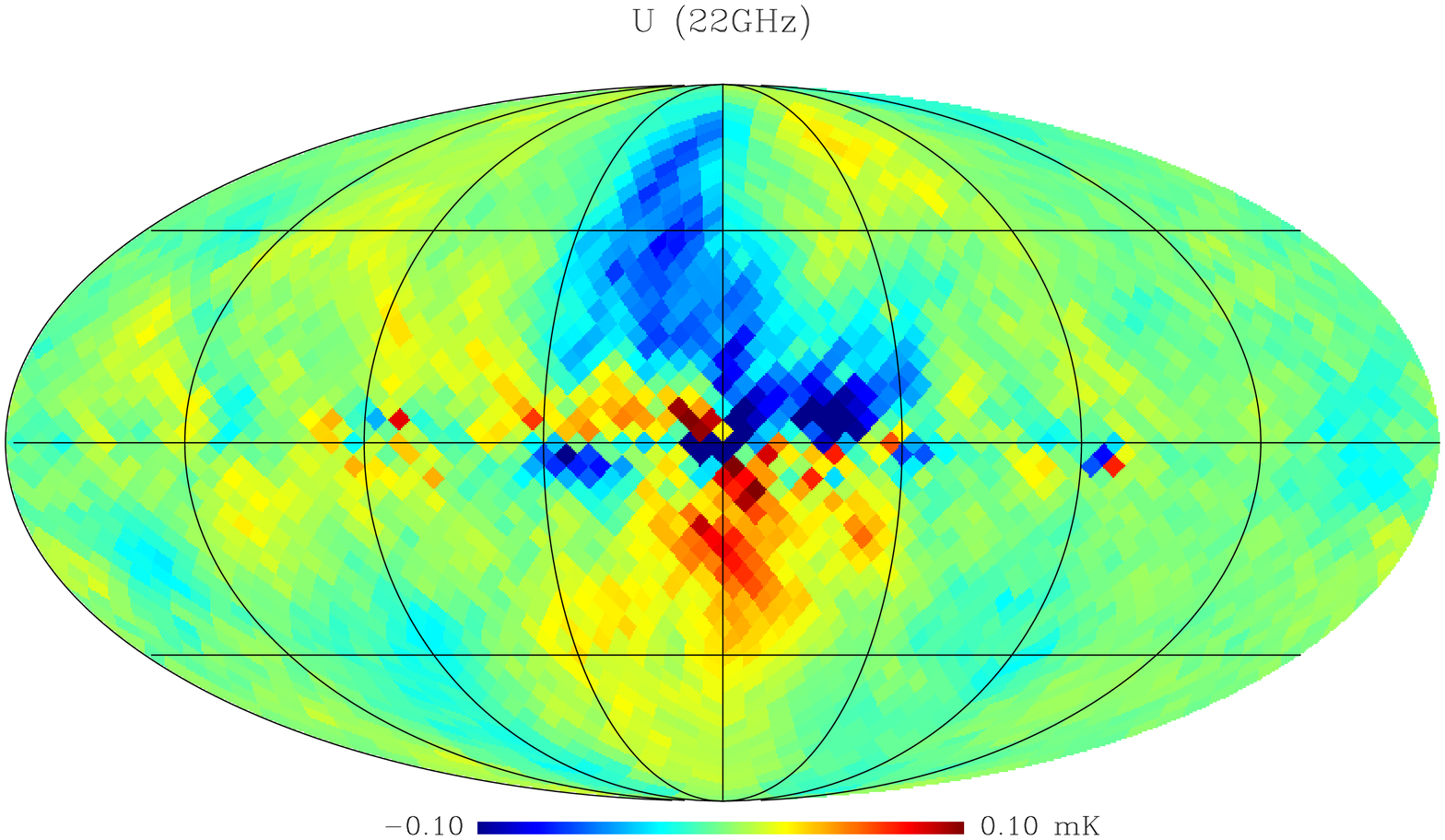} 
\caption{Observed Stokes $Q$ (top) and $U$ (bottom) maps at 22~GHz from WMAP5
  data. The maps are shown in a Mollweide projection of {\sc HEALPix} in
    Galactic coordinates. The center of the map corresponds to $(l,b) =
    (0^\circ,0^\circ)$, and the graticule increases with $\Delta l = \Delta b =
    45^\circ$. The maps are degraded to $\nside = 16$ to enhance the large
  scale pattern. Units are mK (thermodynamic). }
\label{plot_uq}
\end{figure}

Figure~\ref{plot_uq} shows the all-sky Stokes $Q$ and $U$ maps at 22~GHz,
degrading the resolution to enhance the large-scale pattern, using a $\nside =
16$ pixelization (which corresponds to a pixel size of $3.66^\circ$). For all
the computations in this paper, we will use these degraded maps as the input
data.
As shown by the WMAP team \citep{page07}, at this frequency (22~GHz) the
large-scale pattern observed in polarization is completely dominated by the
galactic contribution, and the CMB component is sub-dominant.  Thus, in our
analyses we can safely neglect the contribution of the CMB to the polarization
map.

\begin{figure}
\centering
\includegraphics[height=0.9\columnwidth,angle=90]{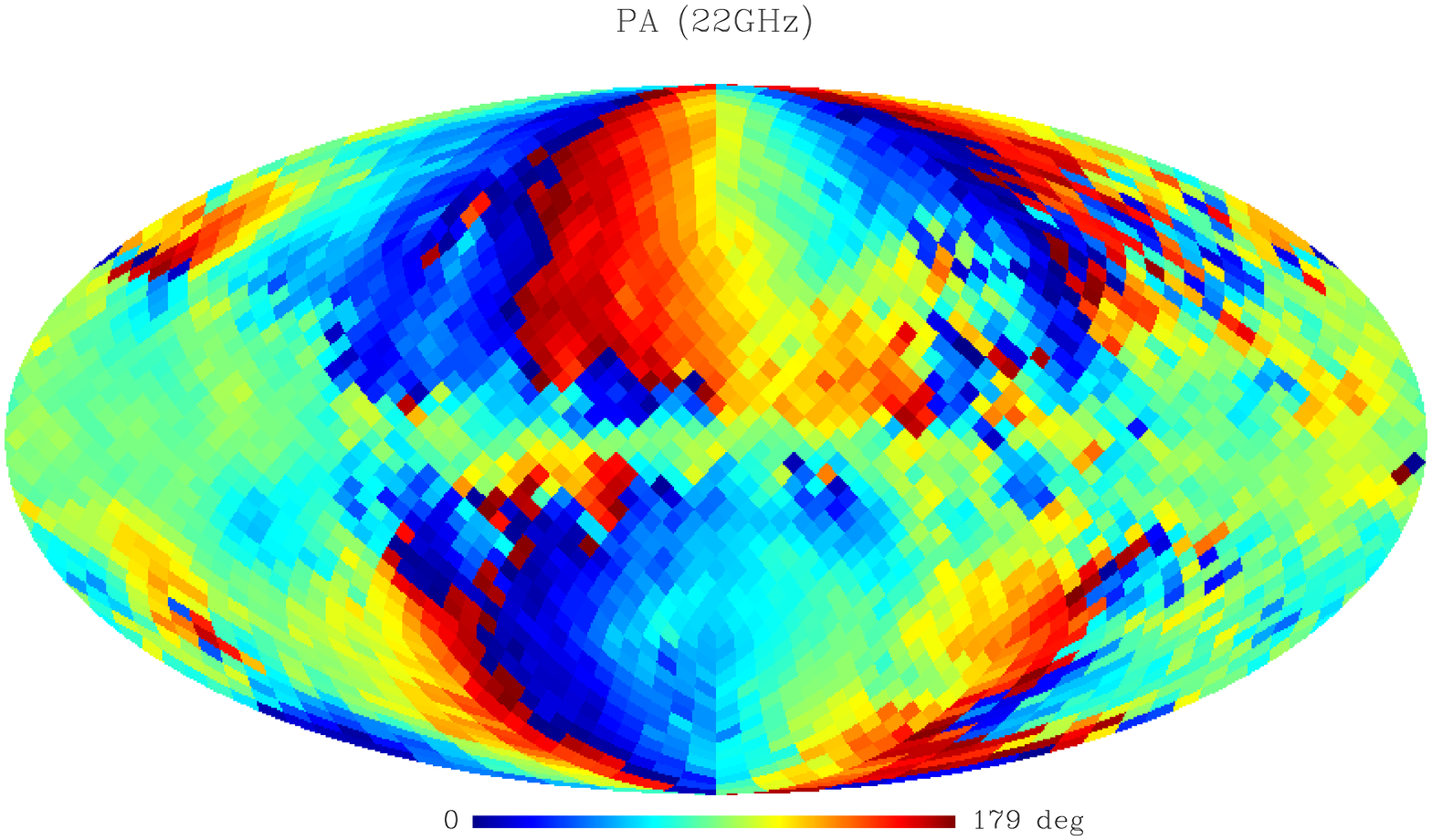}
\caption{Observed direction of the polarization angle (PA) at 22~GHz,
  obtained from the two maps shown in Fig.~\ref{plot_uq}. PA is defined
    here as the local direction of the magnetic field (see equation~\ref{pa}).
  Units are degrees. }
\label{plot_gamma}
\end{figure}

From these two observables (Stokes's $Q$ and $U$ parameters), one can obtain the map of the
direction of the polarization angle (hereinafter, PA) as:
\begin{equation}
\label{pa}
\gamma_{obs}(\hat n) = \frac{1}{2} \arctan \left( \frac{U(\hat n)}{Q(\hat n)} \right) +
\frac{\pi}{2},
\end{equation}
where $\hat n$ is the direction of the line of sight. Note that we
  include in our definition the $\pi/2$ factor, so equation~(\ref{pa}) represents the
  angle of the magnetic field, and takes values in the $[0,\pi]$ region.  Note
that the polarization convention adopted here is the one described in {\sc
  HEALPix}, labelled as COSMO, which differs from the usual IAU convention in a
minus sign for Stokes U parameter. In addition, the WMAP definition of Stokes
parameters includes an additional $1/2$ factor with respect to the definition
used by \cite{chandrasekhar}, which will be the one adopted here.
All these quantities ($U$, $Q$ and PA) are defined in a galactic (heliocentric)
coordinate system. This has to be taken into account when comparing the observed
maps with the models.
Figure~\ref{plot_gamma} shows the observed direction of the PA at the same
$3.66^\circ$ ($\nside = 16$) resolution.


\subsubsection{Noise maps}
\label{sec:noise_maps}
In order to perform the model selection, we also need to estimate the noise maps
associated to the data.
It is important to note that those noise maps should account for the covariances
introduced by all components which are present in the observed data but are not
included in the theoretical model. In particular, it should account for the
instrumental noise component and well as for the random component of the GMF
which is not included in our theoretical model.
The first one could be easily estimated based on the information provided by the
WMAP team about instrument sensitivity and the overall integration time that the
satellite has spent on each particular pixel. However, this would not account
for the second part of the covariance.
Thus, in order to model all the different contributions to the covariance, we
follow a different procedure which makes use of the fact that the original
WMAP-K band maps have a much better angular resolution than the pixel size which
is finally adopted in our analyses.

\paragraph{Noise maps for $Q$ and $U$.}
  In this paper, we have probed three different methods to characterize
  the noise maps associated to $Q$ and $U$. In addition to the pure instrumental
  noise, all these three methods attempt to estimate the contribution to the
  total covariance matrix of the residual astrophysical components which are not
  included in our modelling (for example, the random component of the magnetic
  field, or the variance introduced by point-to-point variations of the mean
  level of the galactic emission). All methods produce a noise map at $\nside
  =16$ resolution, which provides a pixel size of $\sim 3.66^{\circ}$. The three
  methods to obtain the $\sigma_{Q}$ map (or equivalently, the $\sigma_{U}$),
  are:
  \begin{itemize}
  \item Method 1. This corresponds to the same procedure described
    in~\cite{jansson08}. We start from the observed WMAP Stokes-$Q$ map at full
    resolution ($\nside =512$), degrading it to $\sim 0.92^{\circ}$ pixel
    resolution ($\nside =64)$.
    For a given pixel $i$ within our $\nside =16$ pixelization scheme, we obtain
    the associated noise $\sigma_Q(i)$ by computing the square root of the
    variance of the $\sim 0.92^{\circ}$ pixels inside a radius of $2^{\circ}$
    from the center of our pixel $i$.
  \item Method 2. We start from the observed $Q$-map at $nside= 512$, and we
    convolve it with a Gaussian of FWHM$= 1^{\circ}$. For a given pixel $i$
    within our $\nside =16$ pixelization scheme, the associated noise
    $\sigma_Q(i)$ is computed as the square root of the variance in each pixel
    of the smoothed map within a radius of $2^{\circ}$ from the center of our
    pixel $i$.
  \item Method 3. We start from the observed WMAP Stokes-$Q$ map degraded at
    $nside = 16$. For each pixel $i$, the noise $\sigma_Q(i)$ is computed by
    obtaining the square root of the variance inside a circle of radius $r \sim
    7.4^{\circ}$ (i.e. twice the pixel radius). This method produces an unbiased
    estimate of the noise map in the case of uncorrelated noise, provided that
    the scale of variation of the noise map is larger than $7.4^{\circ}$.
  \end{itemize}

\begin{figure}
\centering
\resizebox{\hsize}{!}{\includegraphics[height=0.9\columnwidth,angle=90]{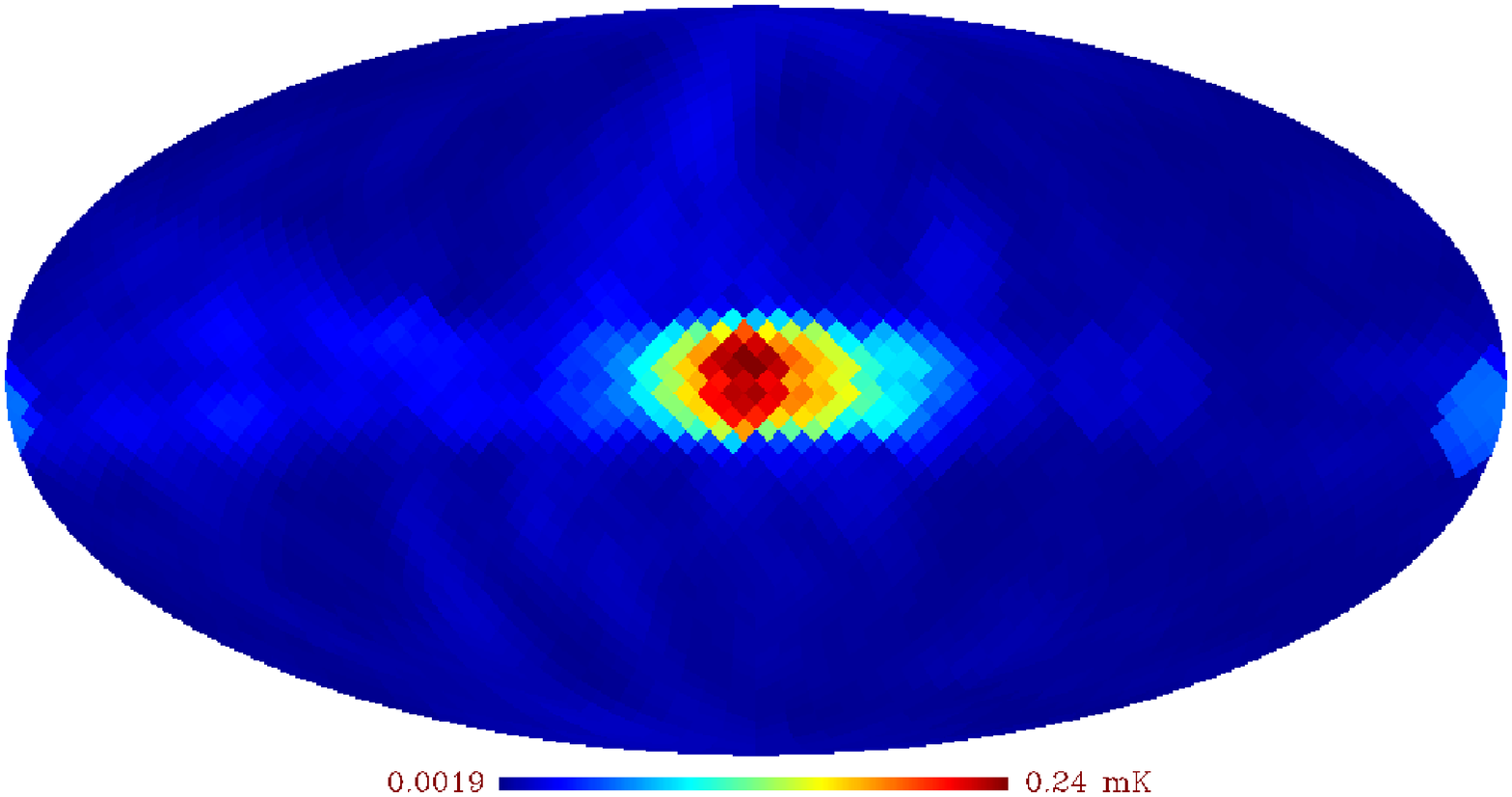}}
\resizebox{\hsize}{!}{\includegraphics[height=0.9\columnwidth,angle=90]{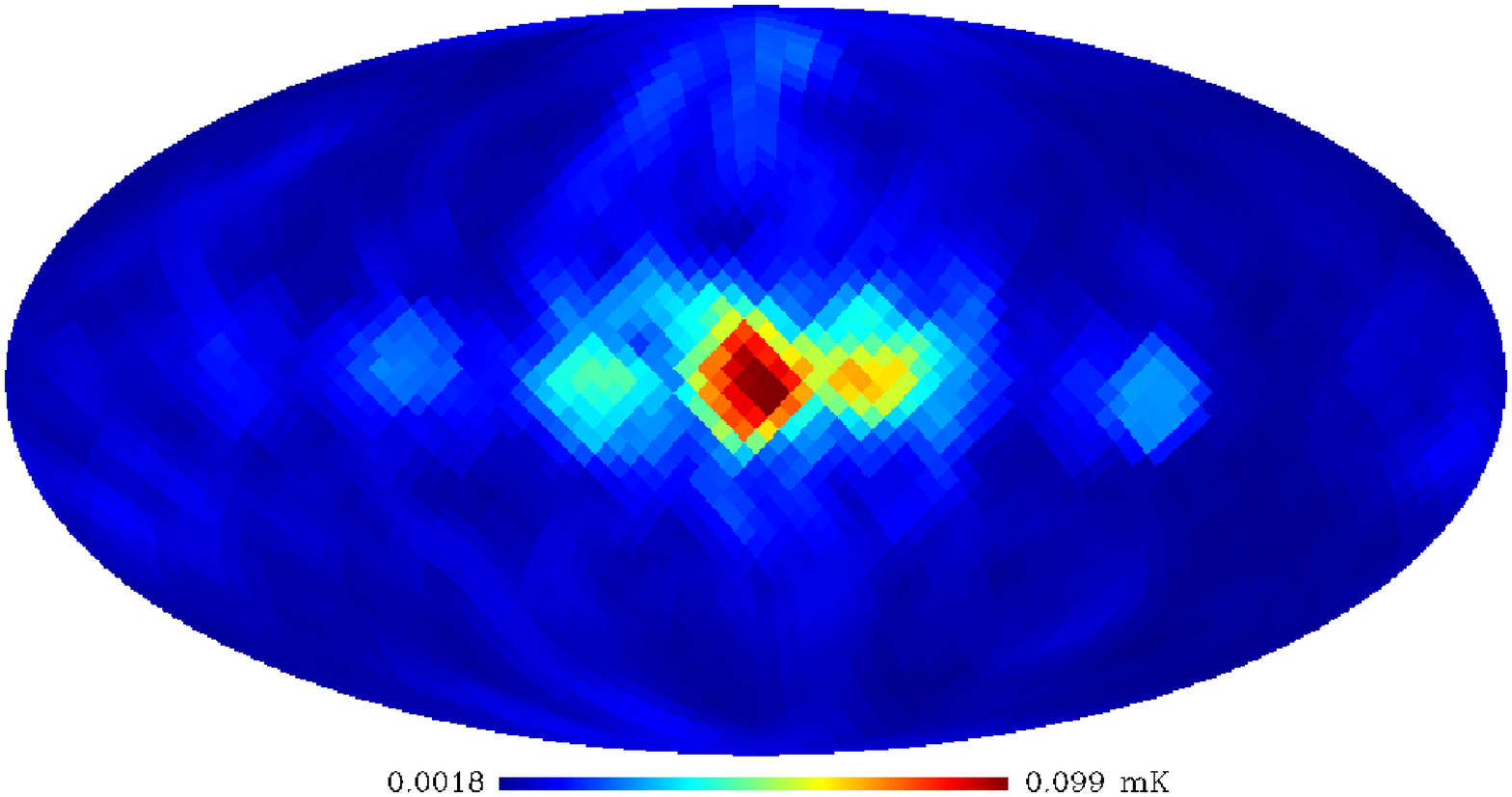}}
\resizebox{\hsize}{!}{\includegraphics[height=0.9\columnwidth,angle=90]{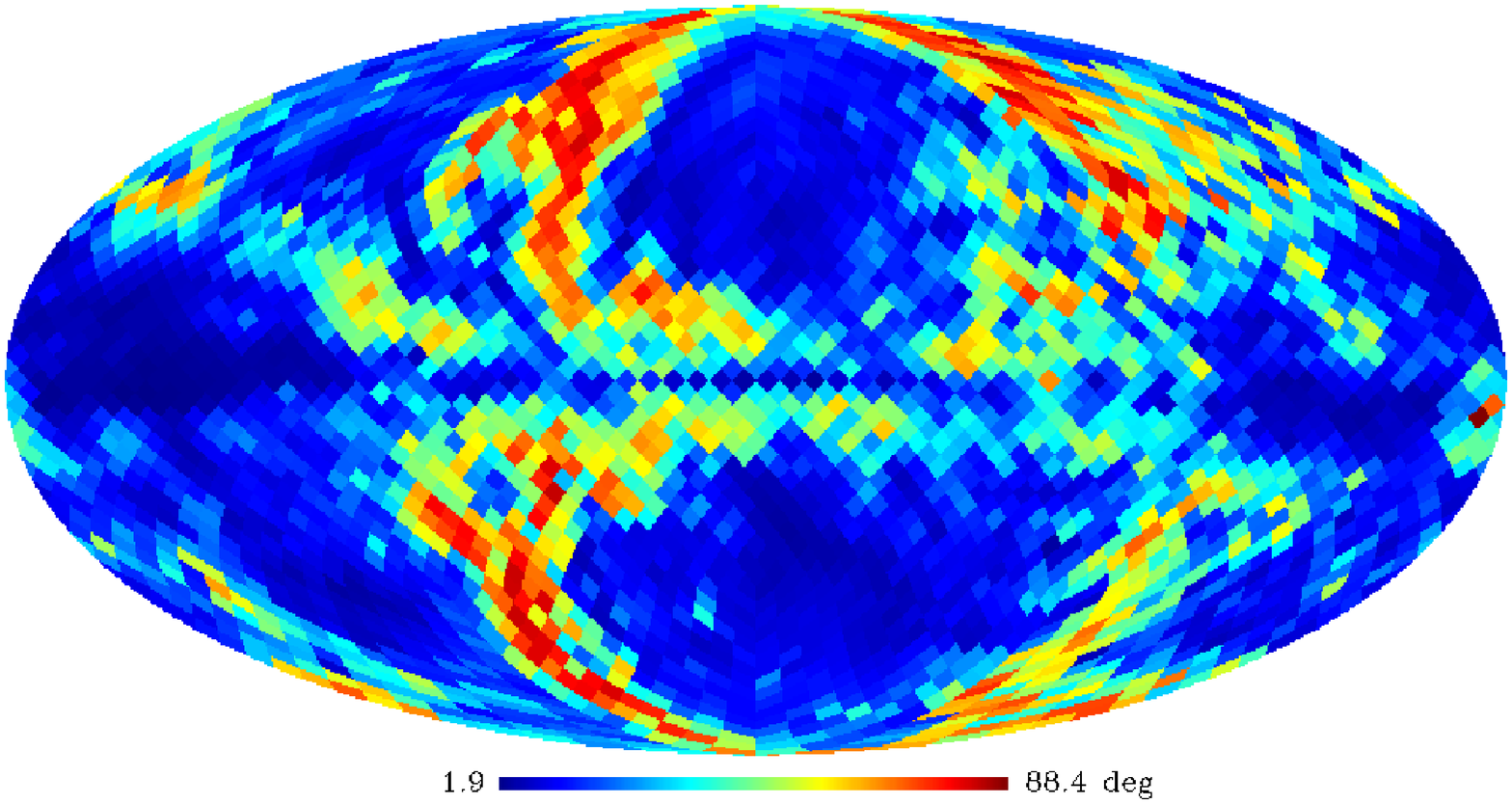}}
\caption{Noise maps for the WMAP5 datasets presented in Fig.~\ref{plot_uq} and
  \ref{plot_gamma}. It is shown the Stokes $Q$ (top), Stokes $U$ (middle) and
  position angle (PA, bottom) maps. Units for Stokes's $Q$ and $U$ maps are mK. Units
  for the PA map are degrees. The noise map for PA has been computed as
  fluctuations around the observed WMAP5 K-band map. }
\label{fig:noise_maps}
\end{figure}

  Each one of these three methods produce different results for the noise map in
  certain areas, specially those dominated by galactic features. In turn, these
  differences imply significant changes in the goodness-of-fit statistic (up to
  a factor of $2-3$ in some cases).
  In general, we find that the larger noise amplitudes are obtained with method
  1, while the smaller amplitudes are obtained with method 2.
  Thus, and as a conservative approach, we decided to present the computations
  of this paper using the method 3.
  The implications of this uncertainty in the determination of the noise maps
  are discussed in section~\ref{sec:conclusions}.
  Noise maps for $(Q,U)$ parameters, obtained with method 3, are shown in
  Figure~\ref{fig:noise_maps}. These maps have been used to obtain the final
  results presented in Tables~\ref{tab:resultsQUhalo}
  and \ref{tab:resultsQUdisco}. 

\paragraph{Noise maps for PA.} 
Once these two noise maps ($\sigma_{Q}$ and $\sigma_{U}$) have been obtained,
one can in principle obtain the noise map associated to the PA
($\sigma_{\gamma}$) from them.  However, there is an important point to stress
here. Equation ~(\ref{pa}), which defines the PA, is not linear in $Q$ and
$U$. This fact implies that the variance map for the PA will depend on the
particular model which is used to compute the average value; or in other words,
the noise map for PA will depend not only on $\sigma_{Q}$ and $\sigma_{U}$, but
also on $Q$ and $U$ themselves.
Therefore, when doing the model selection, the noise map for PA will be also a
function of the model.

For illustration, we compute here the noise map for PA, using as reference model
the observed WMAP-K map. We use here a Monte Carlo method, drawing $N_{\rm sim}$
realizations of pairs of $(Q,U)$ maps, with mean equal to the observed $(Q,U)$
maps, and variance given by $\sigma_{Q}$ and $\sigma_{U}$, as computed in
previous step.
We have checked that using $N_{\rm sim} =5,000$ noise realizations is enough to
get convergence on the $\sigma_{\gamma}$ map at the per cent level.
Figure~\ref{fig:noise_maps} also shows the noise map for the PA, obtained for
this particular case.

%

\subsection{Producing the synchrotron polarized emission for a 
given magnetic field model}

To obtain the predicted polarized synchrotron emission for any of the Galactic
magnetic field models described above, we have developed a code which only
includes the relevant physics, and it is optimised in terms of computational
time with the aim of efficiently perform all the computations.
The code works directly within the {\sc HEALPix} pixelization scheme, and
obtains the predictions for the synchrotron emission directly at the resolution
level that we have chosen (i.e. $\nside =16$).  As discussed above, this
resolution is enough as long as we are interested in the large-scale pattern of
the Galactic emission.

In general, assuming that the cosmic ray spectrum is a power law distribution of
spectral index $p$, we can predict the Stokes's parameters that characterize the
polarization of the synchrotron emission at a certain frequency by computing the
emissivity (energy per unit time per unit volume per frequency per solid angle)
in the two orthogonal directions, parallel and perpendicular to the
  projection of the magnetic field on the plane of the sky. Following the same
as \cite{rybicki}, we have:
\begin{eqnarray}
\nonumber
\epsilon_{\perp}(\nu) &=&  N(r,z) \frac{\sqrt{3}e^{3}}{8\pi m c^{2}} \left(
\frac{4 \pi mc}{3e} \right)^{\frac{1-p}{2}} \nu^{\frac{1-p}{2}}
(B\sin\alpha)^{\frac{p+1}{2}} \qquad \\
& & \Gamma\left ( \frac{p}{4} - \frac{1}{12} \right ) \left [
\frac{2^{\frac{p+1}{2}}}{p+1} \Gamma\left ( \frac{p}{4} + \frac{19}{12} \right )
+ 2^{\frac{p-3}{2}} \Gamma\left ( \frac{p}{4} + \frac{7}{12} \right ) \right ]
\label{eq:power_perp}
\end{eqnarray}
\begin{eqnarray}
\nonumber
\epsilon_{||}(\nu) &=& N(r,z) \frac{\sqrt{3}e^{3}}{8\pi m c^{2}} \left(
\frac{4 \pi mc}{3e} \right)^{\frac{1-p}{2}} \nu^{\frac{1-p}{2}}
(B\sin\alpha)^{\frac{p+1}{2}} \\
& & \Gamma\left ( \frac{p}{4} - \frac{1}{12} \right )
\left [ \frac{2^{\frac{p+1}{2}}}{p+1} \Gamma\left ( \frac{p}{4} + \frac{19}{12} \right )  - 2^{\frac{p-3}{2}} \Gamma\left ( \frac{p}{4} + \frac{7}{12} \right )  \right ]
\label{eq:power_paral}
\end{eqnarray}
where $B$ is the magnetic field, $\nu$ is the frequency, and $e$ and $m$ are the
electron charge and mass, respectively. The function $N(r,z)$ represents the electron number
density at the corresponding position $(r,z)$ in the Galaxy, and it is obtained
from equation~(\ref{ncre}).
From these two equations, the polarized intensity at a given frequency is
obtained by integrating the emissivity along the line of sight:
\begin{equation}
\label{intensity}
I_{\nu}(z,\hat{n}) =
\int{[\epsilon_{\perp}(\nu,z,\hat{n})-\epsilon_{||}(\nu,z,\hat{n})]\exp^{-i
    2\chi(z,\hat{n})}dz}
\end{equation}
where we have set the coordinate system in such a way that z-axis represents the
line-of-sight direction, and the other two directions are contained in
  the plane on sky, with y-axis pointing east and x-axis pointing south
  (i.e. {\sc HEALPix} coordinate convention, as explained in
  Sect.~\ref{sec:wmapdata}).
With these definitions, the Stokes's $Q$ and $U$ parameters are given by
\citep{chandrasekhar}:
\begin{subequations}
\label{eq:q-u-theo}
\begin{align}
\label{eq:q}
Q_{\nu} &= I_{x} - I_{y}\\
\label{eq:u}
U_{\nu} &= 2 \sqrt{I_{x}}\sqrt{I_{y}} \cos{\delta} 
\end{align}
\end{subequations}
where in our case, $\delta = 0^\circ$ (i.e. no desphase is assumed). 

Inserting equation~(\ref{intensity}) into equations~(\ref{eq:q-u-theo}), we obtain
the simulated $Q$ and $U$ components along the line of sight (z-axis) by numerical
integration:
\begin{subequations}
\label{eq:q-u-sim}
\begin{align}
\label{qsim}
Q_{\nu}(\hat{n}) &= K_{Q}(\nu)\int_{LOS}{N_{e}(\hat{n})[B_{x}^{2} - B_{y}^{2}]
  dz} \\
\label{usim}
U_{\nu}(\hat{n}) &= -K_{U}(\nu)\int_{LOS}{ N_{e}(\hat{n}) 2B_{x} B_{y} dz}
\end{align}
\end{subequations}
where we explicitly introduce a minus sign in the equation for the U component,
in order to convert from the IAU convention for the polarization to the {\sc
  HEALPix} convention which is used in the WMAP maps.
The $K_{U}(\nu)$ and $K_{Q}(\nu)$ constants also include the conversion factors
between brightness and temperature.  At 22~GHz, we can safely use the
Rayleigh-Jeans approximation. Substituting the numerical values, we obtain
$K_{Q}(\nu)=1.41\times10^{11}$~mK~cm$^{3}$~($\mu$G)$^{-2}$~kpc$^{-1}$, and
$K_{U}(\nu)=1.25\times10^{11}$~mK~cm$^{3}$~($\mu$G)$^{-2}$~kpc$^{-1}$.
The simulated PA map is derived from these two equations~(\ref{eq:q-u-sim}) as:
\begin{equation}
\label{gammasim}
\gamma(\hat{n}) = 0.5 \arctan \left( \frac{-K_{U}(\nu)\int_{LOS}{ N_{e}(\hat{n})
    2B_{x} B_{y} dz}}{K_{Q}(\nu)\int_{LOS}{N_{e}(\hat{n})[B_{x}^{2} - B_{y}^{2}]
    dz}} \right) + \frac{\pi}{2}
\end{equation}
where $B_{x}$ and $B_{y}$ represent, in our coordinate system, the two
components of the magnetic field which are perpendicular to the line of sight.

Finally, when predicting the expected synchrotron emission for a particular
model, we have included an additional restriction on the line-of-sight
integration, by excluding those points whose galactocentric radial coordinate
$r_{G}$ is smaller than 3~kpc or larger than 20~kpc.
The first restriction excludes the inner region of the Galaxy, where large
deviations from the regular pattern are expected~\citep{larosa06}, while the
second one introduces a radial cut-off. In any case, we have checked that the
results are robust against changes in these numbers.

\subsection{Exploration of the parameter space}

As described in Section~\ref{sec:02}, for each one of the families of GMF
models, we have a set of parameters which define each particular model. Given
that in all cases, the dimension of the parameter space is small (there
are, at the most, four parameters describing a particular model), we decided to
carry out the exploration of the parameter space using a grid-based approach.
For higher dimensions, a Monte Carlo method would be more appropriate.

For each one of the different GMF models, we have constructed three different
grids of models, which we label as ``literature'', ``blind'' and
``non-blind''. The first one is centered around the average values which are
found in the literature for each one of the different parameters. The second one
spans the maximum range which is reasonably expected for each particular
parameter. Finally, the third one is built a-posteriori, once the model
selection has been performed on the previous grid, by centering the new grid
around the best-fit parameters for each case.
Table~\ref{tabla:grid} summarizes all the relevant parameters for each one of
these three grids.  In total, we have computed more than one million models
(290,000 for the blind grid, 970,000 for the non-blind, and
51,000 for the literature) for all the different GMF models described
in Section~\ref{sec:02}.
Each one of those models corresponds to a set of three maps ($Q$, $U$ and PA) of the
expected synchrotron polarized emission of the sky at 22~GHz.

As indication, the average execution time in a standard desktop computer for the
computation of a particular model requires $\la 4$~seconds of CPU time. Thus,
the total CPU time for the construction of all the grids is around 1,500 CPU
hours.

\begin{table*}
\caption[]{Exploration of the parameter space. For each GMF model, we show the
  range of values which has been used to produce three grids (see text for
  details). For each parameter, the three values indicate the minimum, the
  maximum, and the step size (uniform) which was used to build the grid. }
\label{tabla:grid}
\vspace{0.2cm}
\centering
\begin{tabular}{@{} l c c c c r}
\hline \hline
Model &  Parameter   & Blind exploration  & Non-blind exploration  & Literature \\ \hline \hline
LSA   & $\psi_{0}(^{\circ})$   & 10, 80, 2    &  50, 75, 0.5  & 30, 40, 0.2 \\
      & $\psi_{1}(^{\circ})$   &-10, 10, 0.5  & -10, 10, 0.5  & -1.5, 1.5, 0.2\\ 
      & $\chi_{0}(^{\circ})$   &  0, 40, 1    &  15, 50, 0.5  & 15, 30, 0.5\\\hline

ASS(const)& $B_{0}(\mu G)$       & 0.5, 8, 0.5 & 0.2, 9.8, 0.2 &  1, 3, 0.5\\
          & $p(^{\circ})$        &-30, 30, 1   &  0, 40, 0.5   &  -15, 15, 0.5\\
          & $\chi_{0}(^{\circ})$ & 0, 40, 2   &   0, 40, 0.5   &  0, 20, 1    \\\hline

ASS(r) & $r_{1}(kpc)$         & 0, 60, 2      & 0.5, 60.5,1 & 0, 20, 1 \\
       & $p(^{\circ})$        & -30, 30, 1     & 0, 30, 0.5   & -15, 15, 1   \\
       & $\chi_{0}(^{\circ})$ & 0, 40, 2      & 0, 50, 0.5  &  0, 20, 1 \\\hline

BSS$_{\pm}$(const)& $B_{0}(\mu G)$       & 0.5, 8, 0.5 & 0.5, 8.5, 0.5 & 1, 3, 0.5\\
          & $p(^{\circ})$        & -30, 30, 1    & 0, 40, 0.5   & -15, 15, 0.5\\
          & $\chi_{0}(^{\circ})$ & 0, 40, 2     & 0, 35, 0.5     & 0, 40, 2\\\hline

BSS$_{\pm}$(r) & $r_{1}(kpc)$         & 0, 60, 2     & 0.0, 60.0, 1 & 0, 20, 1\\
       & $p(^{\circ})$        & -30, 30, 1    &  0, 30, 0.5   & -15, 15, 1    \\
       & $\chi_{0}(^{\circ})$ & 0, 40, 2     &   0, 35, 0.5    & 0, 20, 1 \\\hline

CCR    & $D_{r}(kpc)$       & 0.1, 10.1, 1  & 1, 11, 0.5  & 0,1.5,0.1 \\ 
       & $w(kpc)$          & 0.1, 19.1, 1  & 3.1, 20.1, 1  & 2,4,0.1 \\
       & $B_{0}(\mu G)$     & 0,9,1       & 2, 10, 0.5  & 1,3,0.2 \\
       & $\chi_{0}(^{\circ})$ & 0, 40, 2     & 4, 50, 1  & 10, 40, 1\\\hline

BT     & $r_{1}(kpc)$     & 0.5, 60.5, 2   &   0, 50, 1   & -  \\
       &$\sigma_{1}(kpc)$ & 0.01, 10.01, 0.5 & 0.01,  5.01, 0.02 & - \\
       &$\sigma_{2}(kpc)$ & 0.01, 10.01, 0.5 & 0.01, 20.01, 0.5 & - \\
\hline
\end{tabular}
\end{table*}

\subsection{Model selection and parameter estimation for each GMF model}

Once we have explored the parameter space with these three grids, we have
derived the best-fit parameters for each one of the GMF models using a bayesian
approach. To this end, we have to both compute the likelihood function
$(\mathcal{L})$, and to provide an expression for the priors.
Once we have obtained the best-fit parameters for each GMF model, different
models with be compared in terms of the reduced $\chi^2$-statistic.

\subsubsection{Likelihood function}
In general, we may assume that the likelihood function is defined as a
multivariate Gaussian when it is written in terms of the observables, i.e.:
\begin{equation}
\ln \mathcal{L} = - \frac{1}{2} \chi^{2}.
\end{equation}
If we assume that the correlations between the different pixels are negligible,
then we have:
\begin{equation}
\chi^{2} = \sum_{i}{\frac{(x_{i}-k_{i})^{2}}{\sigma_{i}^{2}}}
\label{eq:chi2}
\end{equation}
where $x_{i}$ represents the observational data, $k_{i}$ the simulated data and
$\sigma_{i}^2$, the associated noise covariance.
In our case, we have performed two different evaluations of the likelihood.
\begin{itemize}
\item The first one corresponds to a direct comparison of Stokes's $(Q,U)$ parameters. In this case, we have $i=1,...,2N_{pix}$, and $x_i=Q_i$ for
  $i=1,N_{pix}$; and $x_i=U_{i-N_{pix}}$ for $i=N_{pix}+1,...,2N_{pix}$. This
  case will be noted as $\chi^2_{Q,U}$.
\item The second case corresponds to the comparison of PA, so now we have
  $i=1,...,N_{pix}$, and $x_i=\gamma_i$. Note that in this case, $\sigma_i$ will
  depend on $k_i$, and thus the PA noise map needs to be computed for each
  particular model.  This case will be noted as $\chi^2_{PA}$, and will be used
  for comparison with the results of the previous case.
\end{itemize}

Once these two functions ($\chi^2_{Q,U}$ and $\chi^2_{PA}$) are evaluated in all
the data-points of the different grids, the posterior distributions are
obtained, and then marginalised\footnote{The marginal distribution functions are
  obtained by integrating the joint distribution function over the variables
  being discarded. For example, in the case of the LSA model, the marginal
  distribution function for $\mathcal{L}(\psi_0)$ is derived by integrating the
  joint distribution $\mathcal{L}(\psi_0,\psi_1,\chi_0)$ over $\psi_1$ and
  $\chi_0$.}  over all the unwanted parameters.
At the end, we end up with marginal probability distribution functions for each
one of the parameters. From these, confidence intervals are derived as the 0.16,
0.5 and 0.84 points of the cumulative probability distribution function.  Thus,
our parameter estimate is the median of the marginalised posterior probability
distribution function, and the confidence interval encompasses 68 per cent of
the probability.

\subsubsection{Priors}
For the analyses in this paper, we have not introduced prior information on the
parameter values describing any of the models. This is equivalent to say that we
have implicitly adopted a top-hat prior in all the parameters, where the top-hat
function is defined by the parameter ranges presented in
Table~\ref{tabla:grid}. Thus, in all cases, the evaluation of the posterior
would reduce to the computation of the likelihood function $(\mathcal{L})$.

However, and for the case of $(Q,U)$ analysis, we have slightly modified the
standard analysis in the following way. As discussed in Section~\ref{sec:02},
the amplitude of the CR electron spectrum in the solar neighbourhood is highly
uncertain. This will in turn imply a large uncertainty in the recovered strength
for the magnetic field, and moreover, it might produce a bias on the recovered
GMF amplitude.
To account for this additional degree of uncertainty (at least at first order),
we have introduced an additional parameter $\epsilon$, which multiplies to both
the predicted $Q$ and $U$ maps for a given GMF model. Note that such parameter
have no impact on PA. If the CR electron density were entirely correct, then we
would have $\epsilon =1$. If there is an uncertainty in this parameter due to
the modelling of the CR distribution, we account for it by introducing a
Gaussian prior for this additional parameter:
\begin{equation}
-2 \ln \mathcal{L}_{QU} = {\frac{(\epsilon-1)^2}{\sigma_{\epsilon}^{2}}} +\chi^{2}_{Q,U} 
\label{eq:epsilon}
\end{equation}  
and we marginalise over it. Given the existing uncertainty on the CR electron
density, we have chosen a conservative value of $\sigma_{\epsilon} = 0.8$. The
marginalisation over this additional parameter, $\epsilon$, can be done
analytically, yielding:
\begin{equation}
\ln \mathcal{L}_{QU}= \frac{A + 2C}{4B} + \frac{\ln{B}}{4}
\end{equation}
where:
\begin{equation}
A = -\frac{1}{2} \sum_{i}{\frac{x_{i}^{2}}{\sigma_i^{2}} -
  \frac{1}{2\sigma_{\epsilon}^{2}} }
\end{equation}
\begin{equation}
B = \frac{1}{2} \sum_{i}{\frac{k_{i}^{2}}{\sigma_{i}^{2}} +
  \frac{1}{2\sigma_{\epsilon}^{2}} }
\end{equation}
\begin{equation}
C = \sum_{i}{\frac{x_{i}k_{i}}{\sigma_{i}^{2}} +\frac{1}{2\sigma_{\epsilon}^{2}}
}
\end{equation}
Note that this scheme is completely equivalent to the marginalisation over
calibration uncertainties which is also adopted in CMB experiments \citep[see
  e.g.][]{bridle02}.
%

\subsection{Masks}
\label{sec:masks}

Nearby structures in the Galaxy (e.g. supernova remnants) might distort the
regular pattern of the GMF as seen in the synchrotron polarized emission, thus
biasing the determination of some parameters of the GMF model. To probe the
robustness of the estimates as a function of the region which is used for the
fit, we have considered different masks to constrain the global, halo
  and disk components of the GMF. In all cases, we exclude the galactic center
  region defined as the region in which the line of sight crosses a region with
  radius 3~kpc in cylindrical coordinates, that it is not taken into account in
  our integration scheme. To constrain the global component we considered two
  masks, listed in decreasing order of covered sky fraction:

\begin{itemize}
\item Mask 1. It corresponds to the Galactic center region.
\item Mask 2. It combines mask 1 with the well-known local regions with a strong
  polarized intensity that could be interpreted as part of the polarized
  intensity produced by the regular GMF. Here we have considered the four
  ``loops'' described in \cite{berkhuijsen71}, and we have added the polarization
  mask from WMAP team, which includes a better masking of the North Galactic
  spur region and several other small objects (e.g. LMC).
\end{itemize}

To constrain the halo component we exclude the following regions:
\begin{itemize}
\item Mask 3. It excludes the disk defined as the emission contains in $|b| <
  10^{\circ}$.
\item Mask 4. It excludes the emission of the disk (mask 3) and the ``loops''
  defined in mask 2.
\end{itemize}

\noindent
For completeness, we have also considered two masks to constrain the field
pattern in the disk. There are:
\begin{itemize}
\item Mask 5. It excludes the halo region defined as the emission obtained for
  $|b| > 10^{\circ}$.
\item Mask 6. It excludes the halo and the ``loops'' described in
  \cite{berkhuijsen71}.
\end{itemize}

\begin{figure}
\centering
\includegraphics[width=0.9\columnwidth]{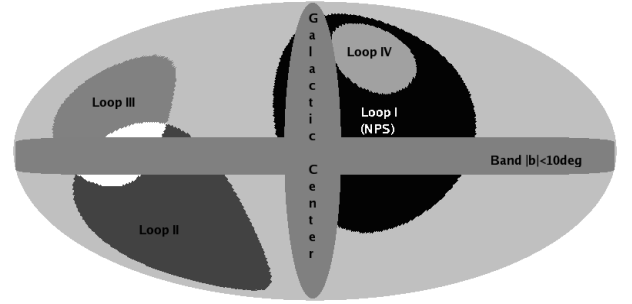}
\caption{Regions used for the definition of the six masks adopted for the
  analyses (see text for details). }
\label{fig:masks}
\end{figure}

\begin{figure}
\centering
\includegraphics[width=0.9\columnwidth]{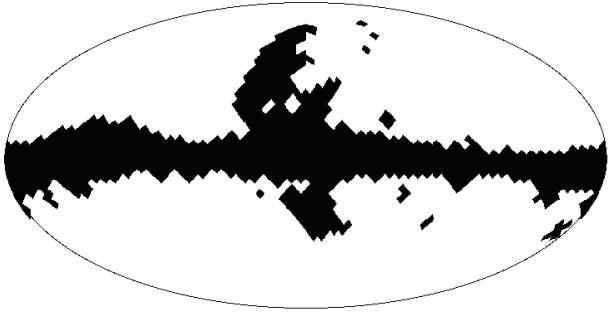}
\caption{Mask for the galactic polarized emission used by the WMAP team.}
\label{fig:mask_wmap}
\end{figure}

In Figure~\ref{fig:masks} we show all regions which have been described in this
subsection. Figure~\ref{fig:mask_wmap} shows the polarization mask used by the
  WMAP team\footnote{http://lambda.gsfc.nasa.gov/product/map/dr3/masks\_get.cfm}.
Table~\ref{tab:masks} presents the detailed information about the sky coverage
of each one of these masks. It is also indicated the number of available pixels
for the analysis (i.e. number of terms in the summation in
equation~(\ref{eq:chi2}). This quantity is relevant in order to compute the
reduced $\chi^2$ for the best-fit models. As a reference, note that in this
pixelization ($\nside=16$), a whole-sky map contains $3072$~pixels. Therefore,
we would have in this case $N_{pix}(QU) = 6144$.

We would like to mention that, even if we use these masks to eliminate the
effects of the random local spurs, other undetected random local features could
be a source of errors in our large-scale models and these errors are difficult
to quantify. But probably our spur masks eliminate the major contribution of
local features.

\begin{table}
\caption[]{Galactic masks used in the analyses. Columns 1, 2 and 3 provide the
  mask identification number and the mask description. Column 4 shows the total
  number of available pixels for the $Q-U$ analysis. Note that for the case of a
  PA analysis, we would have $1/2$ of this value. Last column shows the sky
  fraction available after applying each mask. }
\label{tab:masks}
\vspace{0.2cm}
\centering
\scriptsize{
\begin{tabular}{@{} l c c c c r}
\hline
ID & Mask definition & Region probed & $N_{pix}$ (QU) & $f_{sky}$ \\ 
   &    (regions excluded)    &                        &              & (\%) \\
\hline
1 & GC                           & global & 5432    &  88.4 \\
2 & GC + Loops + mask WMAP       & global &  3452    &  56.2 \\
\hline
3 & GC + Disk                    & halo   &  4416 & 71.9 \\
4 & GC + Disk + Loops + mask WMAP & halo &  2300 &  37.4  \\
\hline
5 & GC + Halo                    & disk  &  1016  & 16.5\\
6 & GC + Halo + Loops            & disk &  482 & 7.8\\
\hline
\end{tabular}
}
\end{table}


\section{Results and discussion}

For each one of the masks described in Table~\ref{tab:masks}, and for each one
of the GMF models presented in Section~\ref{sec:02}, we evaluate the posterior
distribution in each one of the three grids, and marginalising over the relevant
parameters, we obtain the corresponding confidence regions. Our default
  analysis uses the $Q$-$U$ maps, and the noise maps obtained with method 3 in
  section~\ref{sec:noise_maps}. The results are summarized in Tables~\ref{tab:resultsQUhalo}
  and \ref{tab:resultsQUdisco}. 

In order to evaluate which model better reproduces the data, we have used as a
goodness-of-fit the reduced $\chi^2$ statistic, which is obtained as the ratio
of the the minimum value for $\chi^2 (\equiv -2 \ln \mathcal{L}_{QU})$ to the
number of degrees of freedom (hereafter dof). The number of dof is obtained as
$N_{pix}-M$, being $N_{pix}$ the number of terms in Equation~(\ref{eq:chi2}),
and $M$ the number of parameters for the considered GMF model.  Last two columns
in each table present the values for the minimum $\chi^2$, and the reduced
$\chi^2$.

\begin{table*}
  \caption[]{Best-fit parameters for the global and halo component of the GMF
    model (masks 1, 2, 3 and 4), based on the analysis of the
    Stokes's $Q$ and $U$ parameters. For each mask (see Table~\ref{tab:masks}) and
    each GMF model (see section~\ref{sec:02}), it is shown: (a) confidence
    intervals for the parameters listed below derived from the cumulative probability distribution function,
    obtained after marginalising over the rest of parameters; (b) minimum $\chi^2
    \equiv -2 \ln \mathcal{L}_{QU}$ for the ``best-fit model'', defined as the one
    having the smaller $\chi^2$; and (c) minimum $\chi^2$ divided by the number of
    degrees of freedom. Confidence intervals encompass the 68\% of the
    probability, except in those cases in which only upper (lower) limits can be
    given, in which we provide the 95\% limit. For a quick reference, the parameters which
    define each GMF model are: LSA ($\psi_{0}, \psi_{1}, \chi_{0}$); CCR model
    ($D_{r},w,B_{0},\chi_{0}$); ASS ($B_{0},p,\chi_{0}$); BSS$_{\pm}$
    ($B_{0},p,\chi_{0}$); ASS(r) ($r_{1}, p, \chi_{0}$); BSS$_{\pm}$(r)
    ($r_{1},p,\chi_{0}$); BT ($r_{1},\sigma_{1},\sigma_{2}$). Our reference
    computation corresponds to the one labelled as mask 4.}
\label{tab:resultsQUhalo}
\vspace{0.2cm}
\centering
\scriptsize{
\begin{tabular}{@{} l c c c c r}
\hline 
Mask & Model & Confidence regions & Min($\chi^{2}$) for best fit &
Min($\chi^{2}$)/d.o.f.  \\ 
\hline \hline
1 & LSA        &  $23.9^{+0.4}_{-0.3}$, $> 5.8$, $32.4\pm{1.1} $          & 10668.6     & 1.9 \\
1 & CCR        &  $8.5\pm{0.2}$,  $< 5.1$, $< 8.6$, $20.6^{+0.6}_{-0.8} $ & 13828.8 & 2.5 \\
1 & ASS        &  $2.8^{+1.9}_{-0.8}$, $23.9^{+0.3}_{-0.4}$, $20.5^{+0.6}_{-0.8}$        & 10691.1 & 1.9 \\
1 & BSS$_{+}$  & $2.8^{+1.9}_{-0.7}$, $24.5^{+0.5}_{-0.4}$, $< 2.0$    & 12056.9 & 2.2  \\
1 & BSS$_{-}$  &  $2.9^{+1.9}_{-0.8}$, $25.3\pm{0.4}$,  $ < 2.0$  & 11964.6 & 2.2 \\
1 & ASS(r)     &  $ > 38.0$ , $23.8^{+0.5}_{-0.7}$, $21.7\pm{0.7}$    & 10703.0 & 1.9  \\
1 & BSS$_{+}$(r) & $> 50.8$, $24.4\pm{0.6}$, $< 2.0$                & 12103.4 & 2.2 \\
1 & BSS$_{-}$(r)     &  $> 42.4$, $25.1\pm{0.4}$,  $ < 2.0$  & 11984.8 & 2.2 \\
1 & BT         &  $ > 52.5$,  $< 1.0$, $ >1.1  $                           & 15048.3 & 2.8  \\\hline \hline
2 & LSA        &  $23.7\pm{0.5}$, $> 8.1$, $22.0^{+0.8}_{-0.9}  $             & 5456.3 & 1.5  \\
2 & CCR        &  $2.5\pm{0.2}$, $< 5.1$, $< 8.6$, $13.1\pm{0.9} $         & 7053.3 & 2.0 \\
2 & ASS &  $ 2.8^{+1.9}_{-0.8}$, $23.3^{+0.4}_{-0.5}$, $22.4\pm{0.9}$   & 5504.5 & 1.6 \\
2 & BSS$_{+}$ & $2.7^{+1.9}_{-0.7}$, $24.6^{+0.7}_{-0.5}$, $< 2.0$   &  6717.1 & 1.9 \\
2 & BSS$_{-}$ &  $2.7^{+1.9}_{-0.7}$, $24.5^{+0.7}_{-0.5}$, $ < 2.0$  & 6793.8 & 1.9 \\
2 & ASS(r)     &  $> 42.5$,$23.1\pm{0.5}$, $23.8^{+0.8}_{-0.9}$  & 5523.6       & 1.6 \\
2 & BSS$_{+}$(r) &  $> 47.2$, $24.8^{+0.6}_{-0.8}$, $< 2.0$   & 6751.0    & 2.2 \\
2 & BSS$_{-}$(r) &  $> 43.2$, $24.6^{+0.6}_{-0.7}$, $ < 2.0  $ & 6820.0      & 1.9     \\
2 & BT         &  $> 42.0$, $< 1.0$, $0.48\pm{0.2}  $       & 7641.4     & 2.2  \\ \hline \hline
3 & LSA &  $25.5\pm{0.4}$, $< -4.9$, $25.9\pm{0.8}$                    & 8714.8    & 1.9 \\
3 & CCR &  $8.5\pm{0.2}$, $< 5.0$, $< 8.6$, $22.2^{+1.4}_{-1.1}$  & 11735.6   & 2.6 \\
3 & ASS & $2.8^{+2.0}_{-0.7}$, $25.4.0\pm{0.4}$, $25.4^{+0.9}_{-0.8}$   & 8729.8   & 1.9 \\
3 & BSS$_{+}$ & $2.7^{+1.9}_{-0.7}$, $24.6\pm{0.4}$, $< 2.0$   & 9741.2   & 2.2 \\
3 & BSS$_{-}$ & $2.8^{+1.9}_{-0.8}$, $25.4^{+0.4}_{-0.5}$, $< 2.0$   & 9623.0   & 2.1\\
3 & ASS(r)& $2.3\pm{0.6}$, $23.7^{+0.5}_{-0.7}$, $29.4^{+0.8}_{-0.7}$                   & 8533.2   & 1.9 \\
3 & BSS$_{+}$(r) & $20.5^{+11.0}_{-6.0}$, $24.2^{+0.6}_{-0.4}$, $< 2.0$   &  9725.1  & 2.2\\
3 & BSS$_{-}$(r)& $< 2.2$, $24.9^{+0.4}_{-0.6}$, $< 2.0$   & 9478.4  & 2.1\\
3 & BT & $> 33.8$, $2.9^{+0.2}_{-0.3}$, $> 4.7$   & 12099.3   & 2.7\\
\hline \hline
4 & LSA        &  $26.0^{+0.6}_{-0.5}$, $> -6.7$, $32.4^{+1.1}_{-1.2}$ & 3339.0 & 1.4 \\
4 & CCR        &  $3.0\pm{0.2}$, $4.0\pm{0.4}$, $< 8.7$, $20.6^{+1.4}_{-1.1}$ & 4782.9 & 2.1  \\
4 & ASS   &  $2.8^{+2.0}_{-0.8}$, $26.0^{+0.7}_{-0.6}$, $32.5^{+1.2}_{-1.1}$ & 3338.6 & 1.4 \\
4 & BSS$_{+}$ &  $2.6^{+1.8}_{-0.7}$, $23.4\pm{0.7}$, $< 2.0$  & 4164.6    & 1.8\\
4 & BSS$_{-}$ &  $2.5^{+1.8}_{-0.7}$, $23.4^{+0.5}_{-0.7}$, $< 2.0$  & 4163.0  & 1.8 \\
4 & ASS(r)     &  $< 2.5$, $24.3\pm{0.6}$, $30.3^{+1.1}_{-0.9}$             & 3195.7 & 1.3  \\
4 & BSS$_{+}$(r) & $< 2.0$, $24.0\pm{0.6}$, $< 2.0$   &  4037.1  & 1.7\\
4 & BSS$_{-}$(r)     &  $< 2.0$ , $23.8\pm{0.6}$, $< 0.2 $   & 4017.1 & 1.7      \\
4 & BT         &  $> 33.8$, $2.5\pm{0.2}$, $> 3.6  $            & 4824.9 & 2.1      \\ 
\hline \hline
\end{tabular}
}
\end{table*}

\begin{table*}
  \caption[]{The same showed in Table~\ref{tab:resultsQUhalo} for disk 
    component of the GMF model, from the analysis of the
    Stokes's $Q$ and $U$ parameters. Our reference
    computation corresponds to the one labelled as mask 6.}
\label{tab:resultsQUdisco}
\vspace{0.2cm}
\centering
\scriptsize{
\begin{tabular}{@{} l c c c c r}
\hline 
Mask & Model & Confidence regions & Min($\chi^{2}$) for best fit &
Min($\chi^{2}$)/d.o.f.  \\ 
\hline \hline
5 & LSA        &  $16.9^{+1.1}_{-1.2}$, $-1.6^{+4.1}_{-3.8}$, $< 15.9$  &  1729.2  & 1.7 \\
5 & CCR        &  $3.0\pm{0.8}$, $> 14.6$, $< 8.8$, $< 6.6$          &  1970.4  & 1.9\\
5 & ASS        &  $3.1^{+2.0}_{-0.8}$, $17.7\pm{0.8}$, $5.9^{+1.1}_{-1.2}$   & 1664.0  & 1.6 \\
5 & BSS$_{+}$ &  $4.1^{+1.9}_{-1.0}$, $12.2\pm{0.4}$, $18.0^{+1.2}_{-1.3}$  & 1626.1   & 1.6\\
5 & BSS$_{-}$ &  $ 3.9^{+2.0}_{-1.0}$, $13.6\pm{0.4}$, $< 2.0$  &  1775.5  & 1.7 \\
5 & ASS(r)     &  $> 49.1$, $16.8\pm{0.8}$, $6.1\pm{1.2}$        & 1696.7   & 1.6 \\
5 & BSS$_{+}$(r) & $> 46.0$, $11.9\pm{0.4}$, $19.0^{+1.2}_{-1.3}$   & 1647.0   & 1.6\\
5 & BSS$_{-}$(r)     &  $> 49.2 $, $13.0\pm{0.5}$, $< 2.3$   & 1812.8   & 1.8 \\
\hline \hline
6 & LSA        &  $< 19.8$, $> 0.0 $, $17.3^{+1.4}_{-1.3}$  & 672.3    & 1.4 \\
6 & CCR        &  $4.5^{+1.2}_{-0.8}$, $11.4^{+2.9}_{-1.3} $, $< 8.7$, $21.7^{+3.5}_{-4.0}$  & 757.3    & 1.6\\
6 & ASS       &  $3.0^{+2.0}_{-0.8}$, $15.8^{+1.2}_{-1.3}$, $17.3\pm{1.5} $  & 676.8    & 1.4\\
6 & BSS$_{+}$ & $3.9^{+2.0}_{-1.0}$, $10.8\pm{0.5}$, $16.4\pm{2.0}$   &  751.2        & 1.6\\
6 & BSS$_{-}$ &  $3.7^{+2.0}_{-1.0}$, $7.8^{+0.4}_{-0.3}$, $18.0^{+2.6}_{-2.7}$   & 774.2    & 1.6\\
6 & ASS(r)     &  $> 26.0$, $14.6^{+1.3}_{-1.2}$, $18.5^{+1.5}_{-1.6}$         & 680.7    & 1.4 \\
6 & BSS$_{+}$(r) & $> 36.7$, $10.5\pm{0.6}$, $17.6^{+2.0}_{-2.1}$   & 761.2  & 1.6\\
6 & BSS$_{-}$(r)     &  $> 33.9 $, $7.2^{+0.6}_{-0.5}$, $19.2^{+2.9}_{-3.2}$   & 786.5    & 1.6  \\
\hline \hline
\end{tabular}
}
\end{table*}

\subsection{The magnetic field in the Galactic halo}

The results for the halo field are summarized in Table~\ref{tab:resultsQUhalo}.

\begin{figure}
\centering \includegraphics[width=0.9\columnwidth]{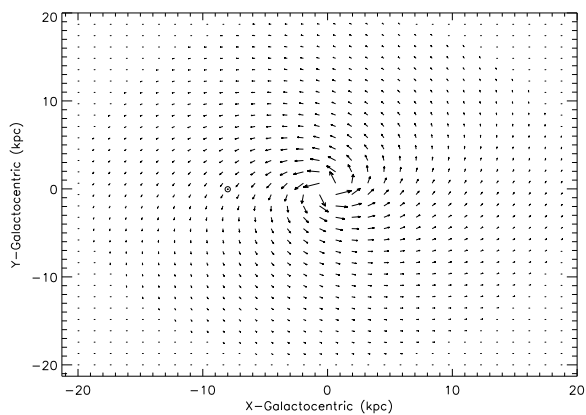}
\caption{Large-scale pattern of the ASS(r) model at $z = 4$~kpc. This
    model provides the best-fit for the halo field. }
\label{fig:gmf-halo}
\end{figure}

\begin{figure}
\includegraphics[width=0.75\columnwidth]{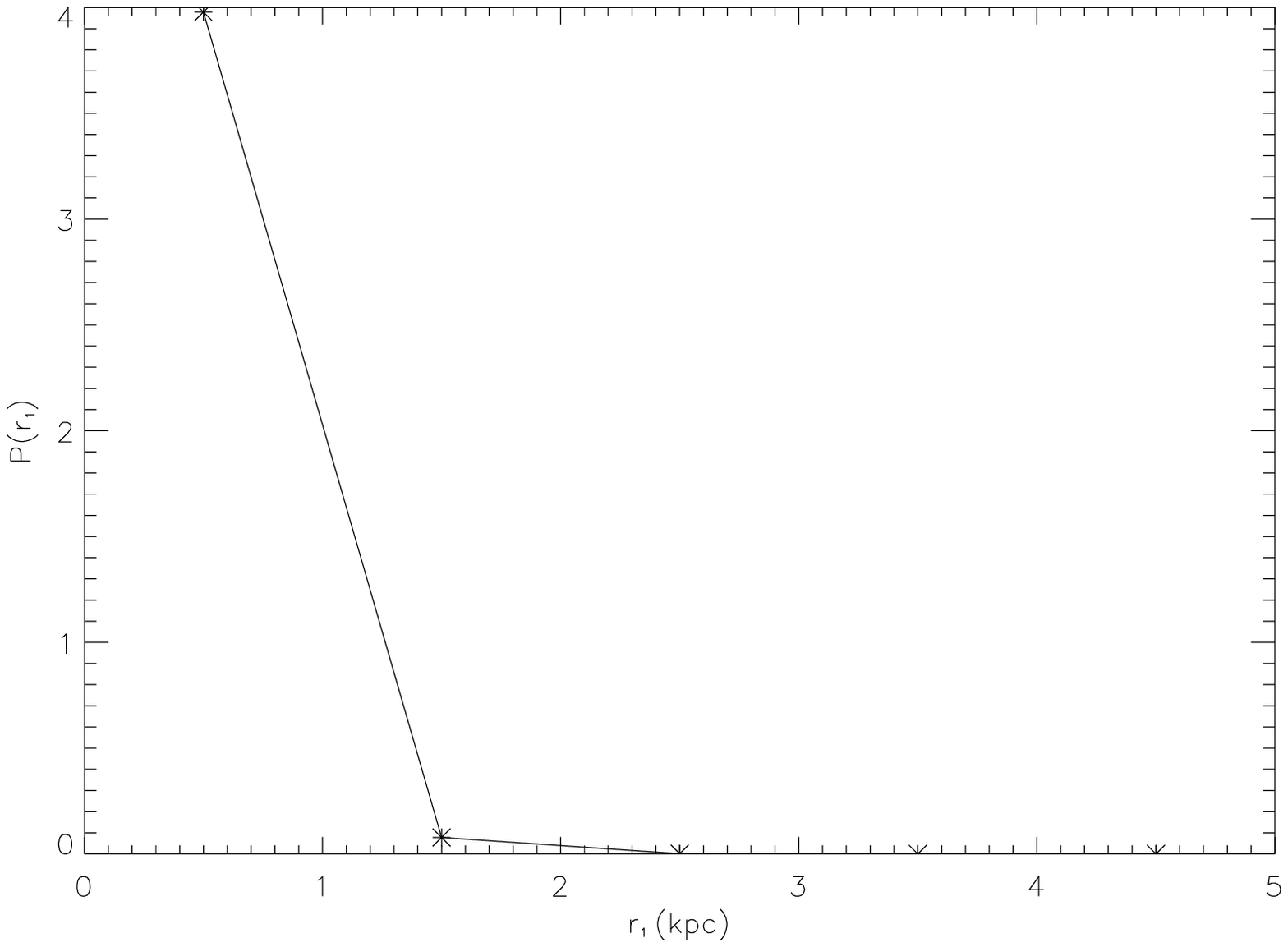}
\includegraphics[width=0.75\columnwidth]{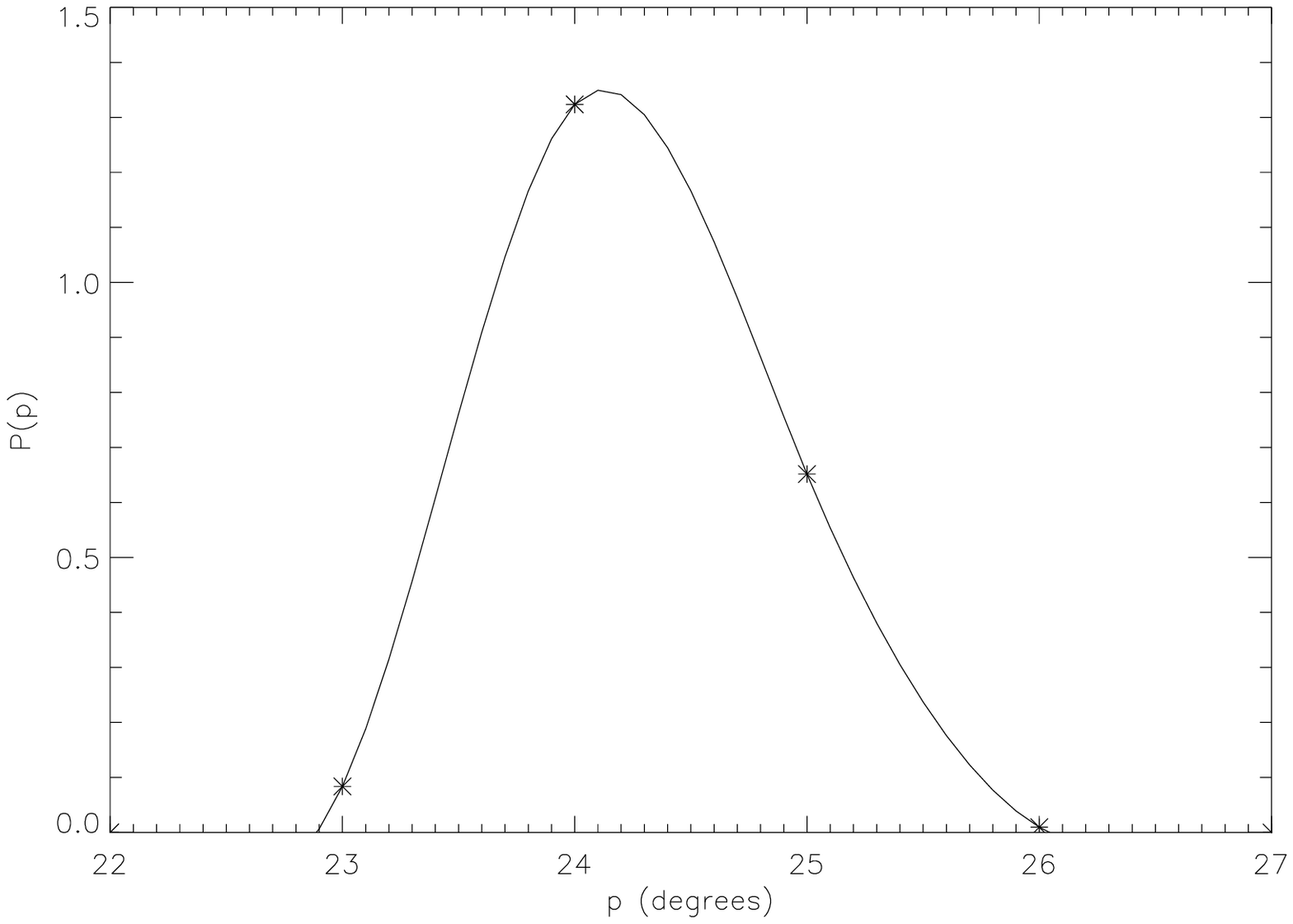}
\includegraphics[width=0.75\columnwidth]{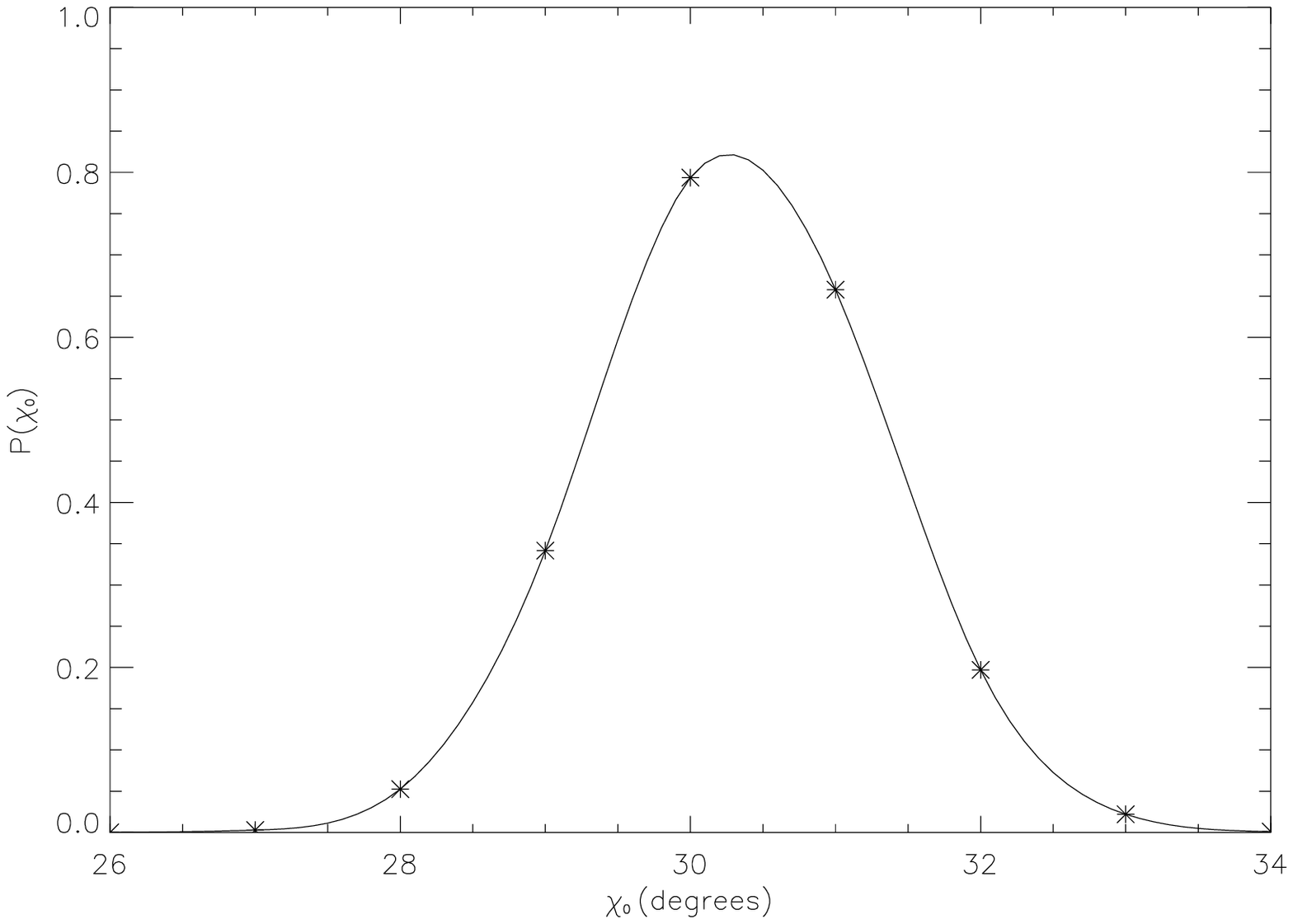}
\caption{One dimensional (marginalised) posterior distribution functions for the
  parameters of ASS(r) halo model (top: $r_1$, middle:$p$ and bottom:$\chi_0$)
  when mask 4 is considered in the QU analysis. }
\label{fig:mag_prob_dist_mask_halo}
\end{figure}

\begin{figure}
  \resizebox{\hsize}{!}{\includegraphics[width=0.2 cm,angle=90]{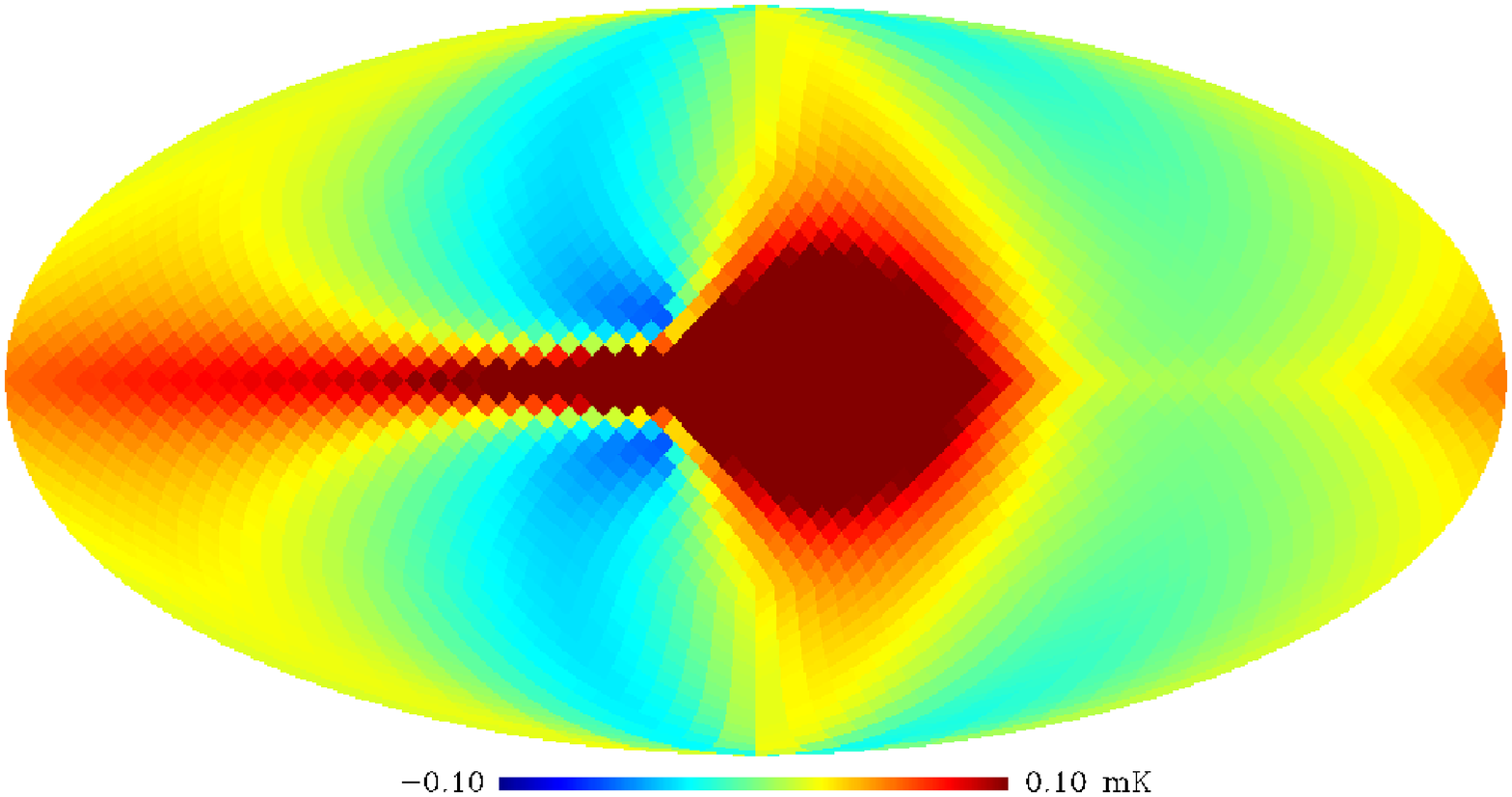}}
  \resizebox{\hsize}{!}{\includegraphics[width=0.2 cm,angle=90]{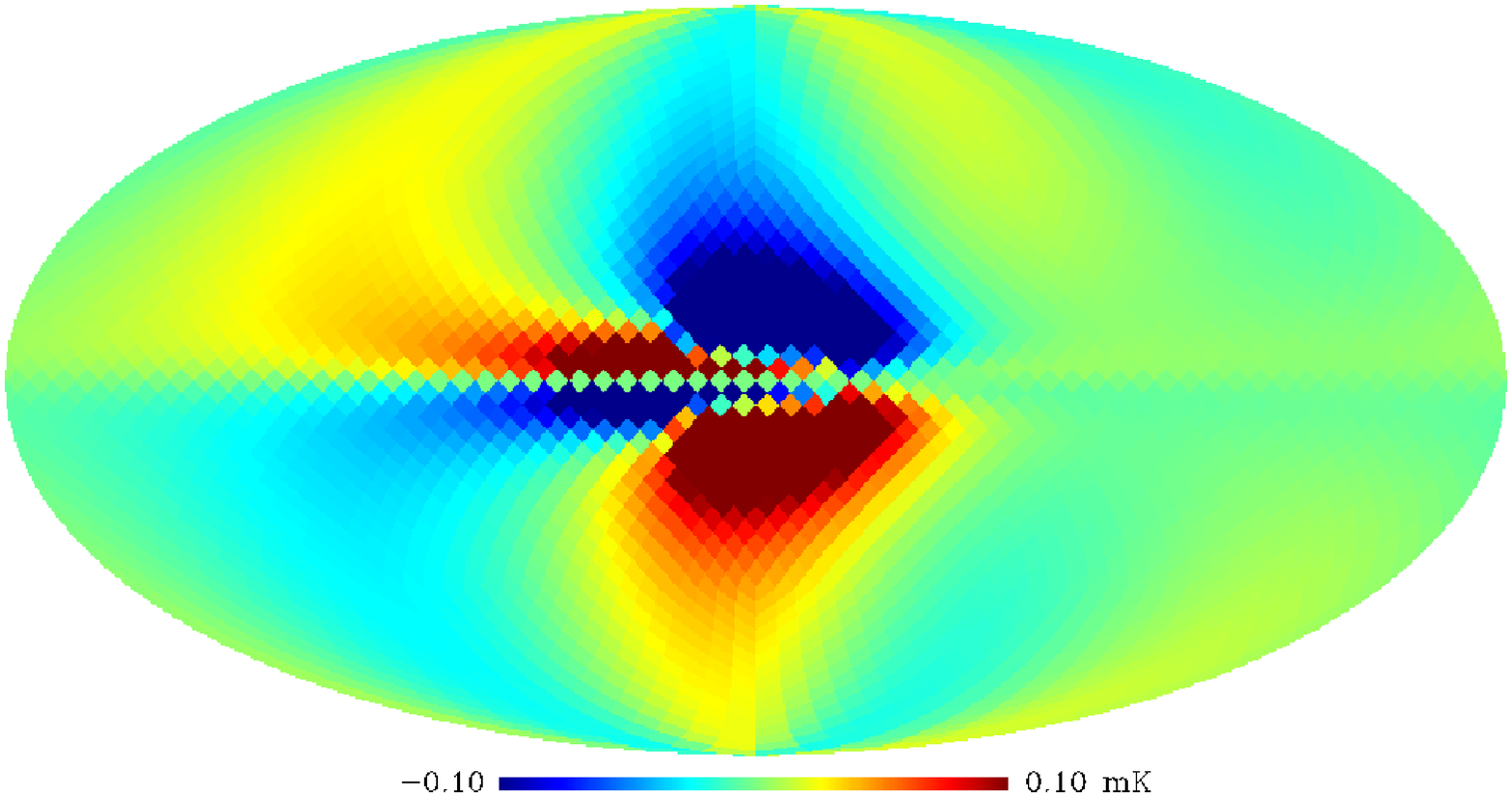}}
  \resizebox{\hsize}{!}{\includegraphics[width=0.2 cm,angle=90]{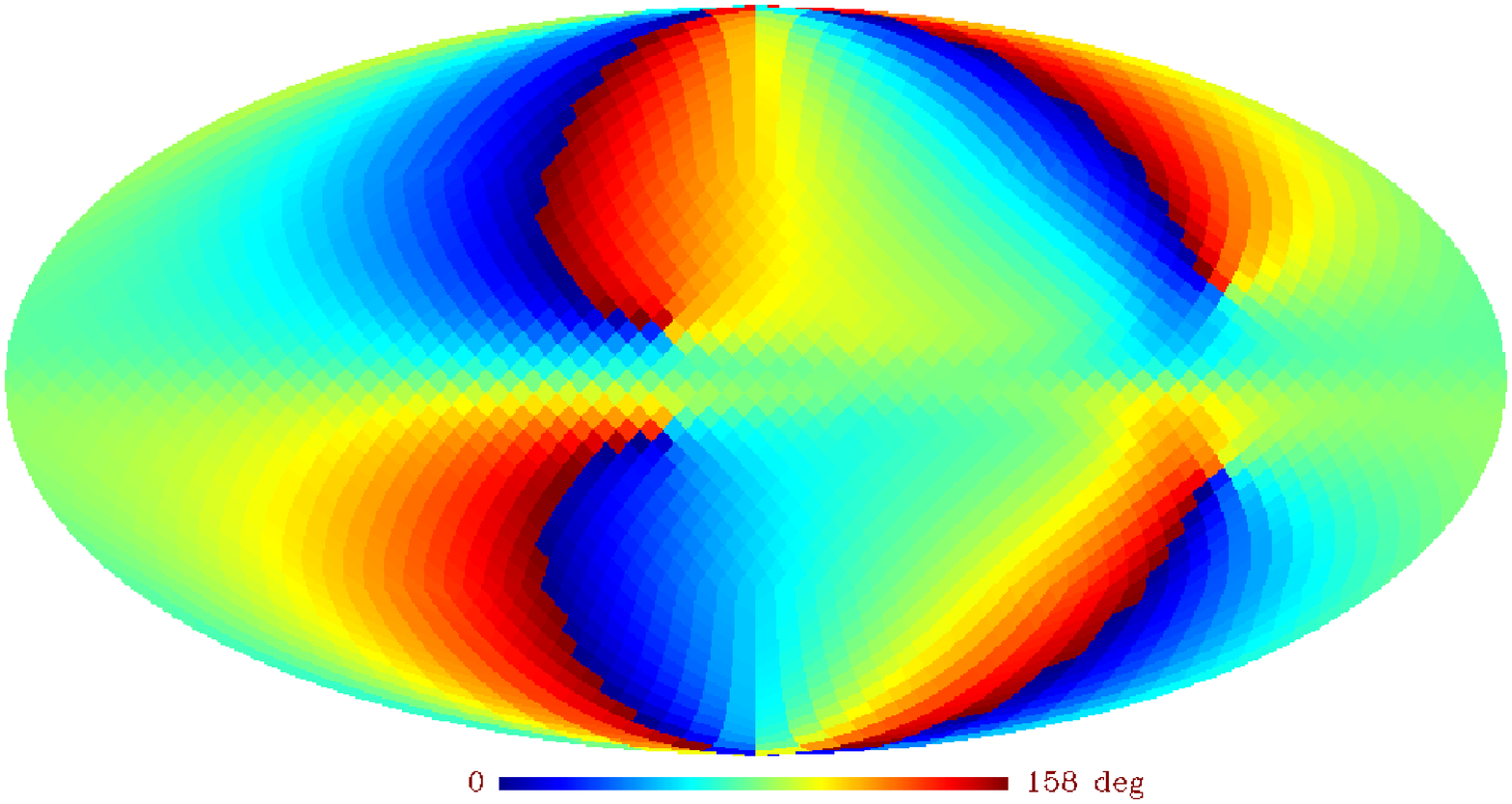}}
\caption{Best-fit for the halo field by using Q,U parameters and PA excluding
  the disk, the galactic center and the ``loops'' (ASS model with radial
  dependence).}
\label{plot_best_fit_mask4}
\end{figure}

Our reference computation corresponds to the case labelled as mask 4, in which
we exclude the emission of the disk, the ``loops'' (including the polarization
mask provided by the WMAP team), and the galactic center region (which is
not accounted for in our analysis).
For this reference case, the model having the minimum reduced $\chi^2$ is the
ASS(r), although the other two axisymmetric models (LSA and ASS) provide
approximately the same goodness-of-fit.
Thus, the large-scale 22~GHz polarized synchrotron emission seems to be more
compatible with some type of axisymmetry, a conclusion also reached
by~\cite{page07}, and also compatible with the results shown by~\cite{sun08}.
For illustration, Figure~\ref{fig:gmf-halo} shows the field pattern of the
best-fit ASS(r) model at $z=4$ kpc; figure~\ref{fig:mag_prob_dist_mask_halo}
shows the marginalised one-dimensional posteriors distributions for the
parameters of this model; and figure~\ref{plot_best_fit_mask4} shows the
  predicted $Q$, $U$ and PA maps for the same best-fit model.
We now discuss separately each one of the relevant parameters.

\paragraph{Radial scale.}
For this ASS(r) model, the derived constraint on $r_1$ is $< 2.5$~kpc (95\%
confidence level). This parameter essentially controls the distance at which the
magnetic field is no longer constant and begins to decrease proportional to
$r^{-1}$ (see Eq.~\ref{eq:radial2}). The obtained value is indeed very small
compared with the radial scale of the electron density or any other scale
distance in our galaxy, suggesting that the data require an important variation
of the field in the inner part of the halo, probably due to the presence of
stronger magnetic fields at the galactic center~\citep[see e.g.][]{roy08}.
Indeed, in the literature, the halo model proposed in~\cite{prouza03} requires
also a small radial scale of $r_{1} = 4$~kpc, being the radial dependence
$\propto \frac{r}{r_{1}} \exp{\left (- \frac{r - r_{1}}{r_{1}} \right )}$ in
that case.

\paragraph{Field strength.}
For the ASS(r), we adopted a fixed values for the magnetic field strength of
$B_{0} = 3$ $\mu$G at the solar neighbourhood. The original LSA model
proposed by \citet{page07} also assumed this fixed value.
However, it considered as a free parameter for the ASS model, and
was found to be in agreement with that value ($B_0 = 2.8^{+2.0}_{-0.8}$~$\mu$G).

\paragraph{Pitch angle.}
The pitch angle was considered in the ASS(r) and ASS models as a constant
free parameter. In both cases, we consistently obtain a value between
$24^{\circ}$ and $26^{\circ}$ respectively. For the LSA model, it is considered
as a radial function with a logarithmic dependence, and in this case the derived
pitch angle at the solar neighbourhood is again $\approx 26^{\circ}$, a value
which is consistent with that given by \cite{page07} of $p = 27^{\circ}$.
As discussed below, these values are larger than the typical pitch angles
obtained for the disk field.

\paragraph{Tilt angle.}
This parameter controls the vertical structure of the field.  The derived tilt
angle in all the axysimmetric models is of the order of $30^{\circ}$, which
implies a vertical component close to 1~$\mu$G at $z = 1$~kpc.
This vertical field component could be identified as the poloidal component
corresponding to the dipole field responsible of the mG vertical component at
the very center~\citep{han94}. It could correspond as well to the vertical
component of the cluster field diffused into the disk by turbulent magnetic
diffusion~\citep[e.g.][]{battaner00}, in this case to the Local Group field.

\paragraph{Other models.}
Table~\ref{tab:resultsQUhalo} show a good consistency between the three
axisymmetric models.
The rest of the families considered here provide slightly worse reduced $\chi^2$
figures, although any of them can not be clearly rejected.
In general, the two families of bisymmetric models provide a slightly poorer
goodness-of-fit. The derived radial scales, field strengths and pitch angles are
in general similar to those found for the axisymmetric models.
However, the tilt values ($\chi_0$) are considerably lower, and in some cases
negligible. This could be due to a compensation produced by the inherent
reversals of the fields for a given direction in a bisymmetric configuration.

The two remaining models (CCR and bi-toroidal) produce the poorer fits
($\chi^{2} = 2.1$), but again they can not be rejected.
Attending to this values, the CCR points to the existence of a reversal at $r
\sim 3$ kpc from the galactic center.
The bi-toroidal model has been suggested as a possible explanation for the halo
double torus that it is interpreted by \cite{han97,han99} and~\cite{han09} as a
consequence of an $\alpha-\Omega$ effect. Another possibility is that the
aforementioned vertical field diffused from the galaxy cluster could be twisted
by differential rotation in the vertical direction, producing toroidal fields
above and below the plane with opposite directions in both
hemispheres~\citep[e.g.][]{battaner00}.

\paragraph{Effect of the loops and the disk emission on the determination of the
  halo field}

As described in section~\ref{sec:masks}, nearby structures in the Galaxy might
introduce biases on the recovered parameters of a given GMF model. Moreover, the
emission of the disk could contaminate the halo field as discussed in
section~\ref{sec01}.
If the disk emission is not excluded by the mask, we could constrain the global
component of the GMF and quantify the impact of the disk emission on the
parameters describing halo field.
The corresponding results are also shown in Table~\ref{tab:resultsQUhalo}.

The influence of disk emission can be seen by comparing the results from masks 2
and 4. In both cases, the best-fits are obtained for axisymmetric models but
when disk emission is not masked out, the reduced-$\chi^{2}$ becomes slightly
poorer. In general, the pitch and tilt angles and the field strength remain
unchanged, with the important exception of the radial scale factor for ASS(r) and
BSS$_{\pm}$(r) models. The inclusion of the disk emission in the analysis
increases drastically this radial scale, probably because as shown below, the
disk does not require strong radial variations.

Finally, we can also evaluate the impact of the loops on the fit by comparing
the results of masks 3 and 4, or 1 and 2.
The basic conclusion in this case is that including the loops regions in the
analysis do not bias significantly the results (even for the radial scale
parameter), but the quality of the fits get worse in all cases.

\subsection{The magnetic field in the disk}
For completeness, in this work we have also used two masks (5 and 6) to study
the magnetic field in the Galactic Plane, by masking the halo emission.
Table~\ref{tab:resultsQUdisco} summarizes the constraints on the different
parameters for this case.
The reference mask now is number 6, which also excludes the
contribution of loops. However, the available sky area for the fit in this case
is very small (7.8\% in total, which corresponds to approximately a 40\% of the
total area of the disk). Because of this limited area, the conclusions on the
magnetic field parameters might be uncertain. Nevertheless, we consider that
still it is important to compare these results with the numbers obtained with
other methods.

Focusing on the mask 6 alone, the three axisymmetric models (LSA, ASS and ASS(r))
provide practically identical value of the goodness-of-fit, which is slightly
better than the other cases.
For these three models, the derived pitch angle values are lower than in the
halo case.
In the solar neighbourhood, the pitch angle of the spiral arm is $\sim
18^{\circ}$ for the stars and $\sim 13^{\circ}$ for all gaseous
components\footnote{Note that these values are translated into our sign
  convention for the pitch angle.} \citep[see][]{vallee95,vallee02}. Therefore,
our best-fit suggests that the magnetic arms follow the gas structure ($p \sim
14^{\circ}-15^{\circ}$).
However, we note that \cite{jansson09} found a value of $p\approx 35^{\circ}$
for the ASS+RING model proposed by~\cite{sun08}, which is not compatible with the
one obtained here.

\begin{figure}
\centering
\includegraphics[width=0.9\columnwidth]{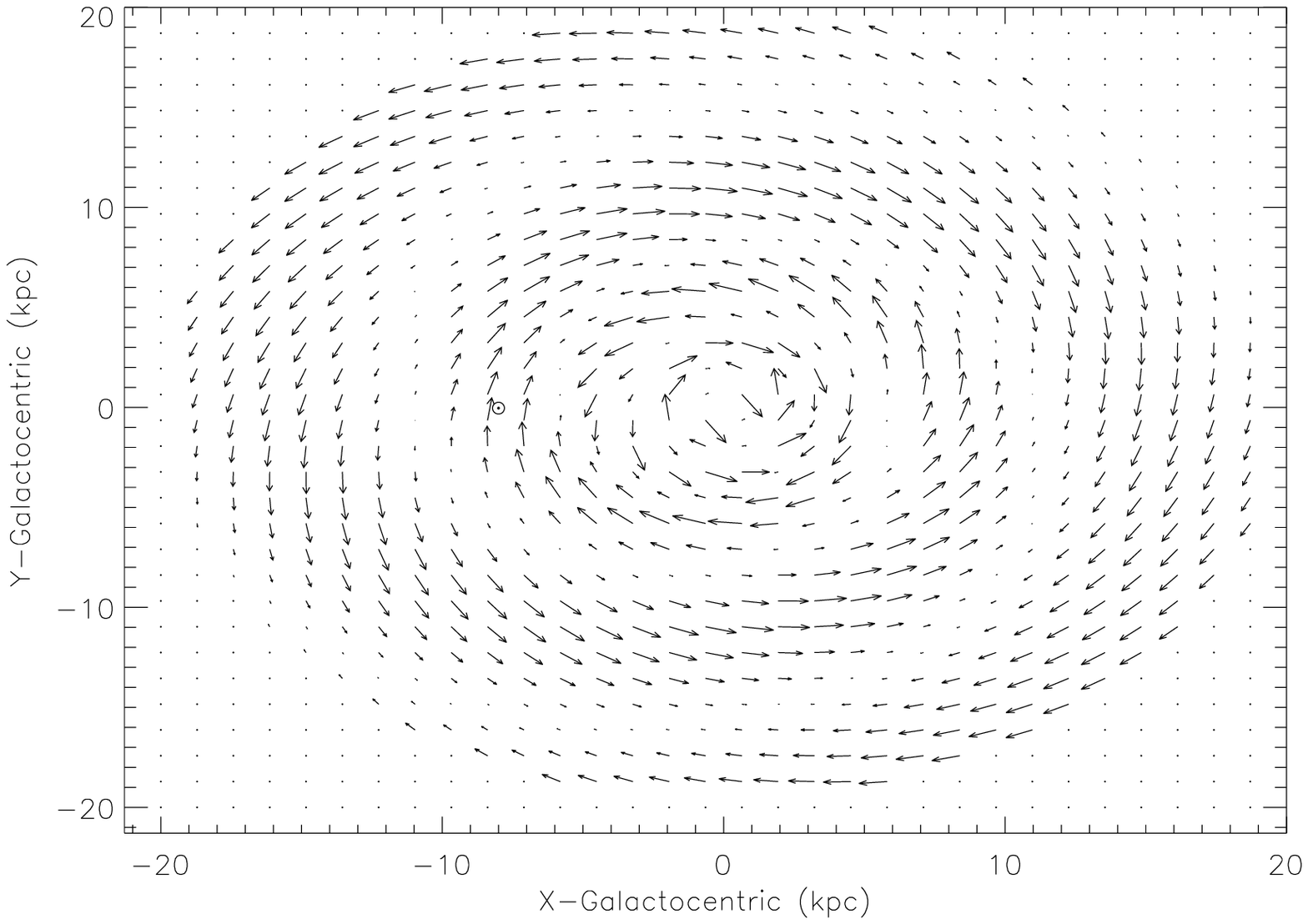}
\caption{Large-scale pattern of the best-fit BSS$_{+}$ model for the disk
  emission in the galactic disk ($z=0$).}
\label{fig:plot_bss}
\end{figure}

Among the two families of bisymmetric models, both of them provide good results,
and again, the pitch angle values are low. Moreover, the derived constraints
(between $7^\circ-10^\circ$) are fully compatible with the results obtained by
other authors.
In the literature, values also range from $7.2^{\circ}$ to $11^{\circ}$ (see
\citet{han94} and \citet{han01} for values obtained with Faraday rotation of
pulsars; and \citet{heiles96}, for values obtained with polarized starlight).
For comparison purposes, Figure~\ref{fig:plot_bss} shows the pattern of our
best-fit BSS$_+$ model in the galactic disk ($z=0$), which is similar in shape
to the results obtained by other authors (see e.g. figure~5 in \cite{han94}).

We note that the values derived for the tilt angle in all models are of the
order of $\la 19^\circ$, which again imply a $z$-dependence of the field
strength within the disk which is compatible with those values observed by
\cite{han94} and up to $0.4 \mu$G for the thin disk.


Finally, we would like to mention that we do not expect changes in these results
if we include a more refined treatment of the turbulent magnetic field in the
analysis, since the polarized synchrotron emission comes from the regular
pattern of the GMF which is located in the inter-arms regions
\citep{beck07proc}. Indeed, we find that the results obtained with the other two
methods for the noise determination (see Sect.~\ref{sec:noise_maps}) are fully
consistent with those presented here for all masks and all parameters.  We only
found a small dependence of the constrained value for $\chi_0$ for axisymmetric
and bisymmetric models with the three noise maps, but which is of the order of
$\la 15$ per cent.

%

\subsection{Comparison of the result with the PA analysis}

For comparison purposes, in this paper we also have done the analyses
  using the PA information alone. As discussed elsewhere, such analysis provides
  a limited amount of information, due to the fact that the PA is not sensitive
  to some parameters, as a constant field strength.

  In general, the results using the PA provide compatible results to those of
  the QU analysis, although there is a larger number of unconstrained parameters
  and a poorer goodness-of-fit.  As illustration, the ASS(r) analysis with mask 4
  gives $ r_1 < 45.9$~kpc, $p \sim 20.0^{\circ}$, and $\chi_0 \sim 21.0^{\circ}$
  ($\chi^{2} \sim 2.3$); while for the ASS case it gives $p \sim 20.5^{\circ}$,
  $\chi_0 \sim 19.0^{\circ} $ and no constraint on $B_0$ ($\chi^{2} \sim
  2.3$). The pitch and tilt angles are compatible with those derived from the QU
  analysis. The best fit is given by the LSA with $\psi_{0} > 22.3^{\circ}$,
  $\psi_{1} > 8.6^{\circ}$ and $\chi_{0} < 18.1^{\circ}$ ($\chi^{2} \sim
  2.3$).

%
\section{Conclusions}
\label{sec:conclusions}
In this work, we have constrained the regular galactic magnetic field by
  using the polarized emission at 22 GHz. To this aim, we have considered
nine models of the galactic magnetic field, each defined by
three or four free parameters and for six different masks to interpret the
polarized maps from WMAP5. The combination of models, free parameters, and masks
produce a very large number of simulated maps to be compared with the
observational ones in Stokes's Q and U parameters, which in turn provide
valuable constraints to determine the three dimensional configuration
of the magnetic field of our galaxy.


The family of GMF models that better describes the halo emission is the
axisymmetric one, although any of the other considered models can be rejected
based on their goodness-of-fit.
The magnetic spiral arms have a pitch angle of $p \approx 24^{\circ}$, and a
tilt angle of $\chi_0 \approx 30^{\circ}$, implying a strong vertical field of
$\sim 1$ $\mu$G at $z = 1$ kpc.
When a radial variation is fitted, the models generally require a fast variation
in the inner part of the galaxy ($r_{1} < 2.5$~kpc).
%

We would like to stress that an accurate determination of the covariance matrix
which accounts for both for the noise and the residual astrophysical components
is very complicated. In this work, we have explored in detail three different
methods, and found that they may lead to differences of a factor of $\sim 2-3$
in the goodness-of-fit values, while the values of the best-fit parameters for
each model do not vary significantly.  In practice, this means that rejecting a
model based on the goodness-of-fit could be inappropriate, although the relative
differences of the $\chi^2$ statistic between models can be used for a
comparison and for selecting preferred models.

We have also tried to constrain the disk field parameters using the polarized
synchrotron emission, despite of the fact that the fitted region is very small
($\sim 8\%$ of the sky).
Here, the obtained results a remarkably consistent with those obtained
with other methods. 
In this case, all the considered models give a very similar goodness-of-fit,
with a (very small) preference for an axisymmetric model.
The data does not require a radial dependence of the strength, and the pitch
angle values are much smaller than in the halo case.
Indeed, the constrained pitch angles are compatible with those found by using
other observational methods as Faraday rotation of pulsars~\citep{han06}
However, we note that this conclusion has not been reached by recent
results where polarized synchrotron emission is used \citep[see
e.g.][]{jansson09}.
The tilt angle in the disk is $\chi_0 \sim 17^{\circ}$, which implies a vertical
field structure being $B_{z} \sim 0.1$ $\mu$G at $z = 200$~pc.  This value is
compatible with that found by \citet{han94}, based on the rotation measure of
pulsars.

We remark that there are still some important uncertainties in the modelling of
the synchrotron emission. In particular, probably we need a better knowledge of
the distribution of cosmic rays, which is today a clear source of
indeterminations in this type of analyses.
The detailed modelling of the halo field could have an influence on the
cosmic rays trajectories and could be crucial for the direct detection for the
primordial magnetic fields.

Finally, we expect that the higher sensitivity and angular resolution in the
polarized channels of the PLANCK telescope \citep{planck06,tauber2010}, the low
frequency channels of the QUIJOTE-CMB experiment \citep{rubino-martin10} and
experiments as LOFAR and SKA \citep{beck09proc}, will provide a much larger
improvement of our knowledge of the galactic magnetic field.


\begin{acknowledgements}
  We acknowledge Dr.~Marco Tucci and Dr.~Juan Betancort-Rijo for useful
  discussions and Dr.~\'Angel de Vicente for computing support. We acknowledge
  the MPIfR 'magnetic group' and specially Prof. Richard Wielebinski for help
  and useful discussions.
  We acknowledge the use of the Legacy Archive for Microwave Background Data
  Analysis (LAMBDA). Support for LAMBDA is provided by the NASA Office of Space
  Science.
  Some of the results in this paper have been derived using the {\sc HEALPix}
  \cite{gorski05} package.
  To reduce the computational time needed for the production of the polarized
  synchrotron maps in this paper, we made use of the Condor workload management
  system (http://www.cs.wisc.edu/condor/) installed at the IAC.
  This project has been partially supported by Spanish MEC Grant ESP
  2004-06870-C02.
  This work has been partially funded by project AYA2007-68058-C03-01 of the
  Spanish Ministry of Science and Innovation (MICINN). JAR-M is a Ram\'on y
  Cajal fellow of the MICINN.
\end{acknowledgements}

%


\begin{thebibliography}{55}
\expandafter\ifx\csname natexlab\endcsname\relax\def\natexlab#1{#1}\fi

\bibitem[{{Abdo} {et~al.}(2009){Abdo}, {Allen}, {Aune}, {Berley}, {Casanova},
    {Chen}, {Dingus}, {Ellsworth}, {Fleysher}, {Fleysher}, {Gonzalez},
    {Goodman}, {Hoffman}, {Hopper}, {H{\"u}ntemeyer}, {Kolterman}, {Lansdell},
    {Linnemann}, {McEnery}, {Mincer}, {Nemethy}, {Noyes}, {Pretz}, {Ryan},
    {Parkinson}, {Shoup}, {Sinnis}, {Smith}, {Sullivan}, {Vasileiou}, {Walker},
    {Williams}, \& {Yodh}}]{abdo09}{Abdo}, A.~A., {Allen}, B.~T., {Aune}, T.,
  {et~al.} 2009, \apj, 698, 2121

\bibitem[{{Battaner} {et~al.}(2009){Battaner}, {Castellano}, \&
    {Masip}}]{battaner09let} {Battaner}, E., {Castellano}, J., \& {Masip},
  M. 2009, \apjl, 703, L90

\bibitem[{{Battaner} \& {Florido}(1995)}]{battaner95} {Battaner}, E. \&
  {Florido}, E. 1995, \mnras, 277, 1129

\bibitem[{{Battaner} \& {Florido}(2000)}]{battaner00} {Battaner}, E. \&
  {Florido}, E. 2000, Fundamentals of Cosmic Physics, 21, 1

\bibitem[{{Battaner} \& {Florido}(2007)}]{battaner07} {Battaner}, E. \&
  {Florido}, E. 2007, Astronomische Nachrichten, 328, 92

\bibitem[{{Battaner} \& {Florido}(2009)}]{battaner09} {Battaner}, E. \&
  {Florido}, E. 2009, in IAU Symposium, Vol. 259, IAU Symposium, 529--538

\bibitem[{{Battaner} {et~al.}(1992){Battaner}, {Garrido}, {Membrado}, \&
    {Florido}}]{battaner92} {Battaner}, E., {Garrido}, J.~L., {Membrado}, M., \&
  {Florido}, E. 1992, \nat, 360, 652

\bibitem[{{Beck}(2009)}]{beck09} {Beck}, R.\ 2009, IAU Symposium, 259, 3

\bibitem[Beck(2009)]{beck09proc} Beck, R.\ 2009, Revista Mexicana de Astronomia
  y Astrofisica Conference Series, 36, 1

\bibitem[{{Beck}(2007)}]{beck07proc} {Beck}, R. 2007, in EAS Publications
  Series, Vol.~23, EAS Publications Series, ed. M.-A. {Miville-Desch{\^e}nes} \&
  F.~{Boulanger}, 19--36

\bibitem[{{Beck} {et~al.}(1996){Beck}, {Brandenburg}, {Moss}, {Shukurov}, \&
    {Sokoloff}}]{beck96} {Beck}, R., {Brandenburg}, A., {Moss}, D., {Shukurov},
  A., \& {Sokoloff}, D.  1996, \araa, 34, 155

\bibitem[{{Berkhuijsen} {et~al.}(1971){Berkhuijsen}, {Haslam}, \&
    {Salter}}]{berkhuijsen71} {Berkhuijsen}, E.~M., {Haslam}, C.~G.~T., \&
  {Salter}, C.~J. 1971, \aap, 14, 252

\bibitem[{{Beuermann} {et~al.}(1985){Beuermann}, {Kanbach}, \&
    {Berkhuijsen}}]{beuermann85} {Beuermann}, K., {Kanbach}, G., \&
  {Berkhuijsen}, E.~M. 1985, \aap, 153, 17

\bibitem[{{Bluemer} \& {for the Pierre Auger Collaboration}(2008)}]{auger08}
  {Bluemer}, J. \& {for the Pierre Auger Collaboration}. 2008, ArXiv e-prints

\bibitem[{{Bridle} {et~al.}(2002){Bridle}, {Crittenden}, {Melchiorri}, {Hobson},
    {Kneissl}, \& {Lasenby}}]{bridle02} {Bridle}, S.~L., {Crittenden}, R.,
  {Melchiorri}, A., {et~al.} 2002, \mnras, 335, 1193

\bibitem[{{Brown} {et~al.}(2007){Brown}, {Haverkorn}, {Gaensler}, {Taylor},
    {Bizunok}, {McClure-Griffiths}, {Dickey}, \& {Green}}]{brown07} {Brown},
  J.~C., {Haverkorn}, M., {Gaensler}, B.~M., {et~al.} 2007, \apj, 663, 258

\bibitem[Carretti et al.(2008)]{carretti08} Carretti, E., McConnell, D.,
  Haverkorn, M., Bernardi, G., McClure-Griffiths, N.~M., Cortiglioni, S., \&
  Poppi, S.\ 2008, Mapping the Galaxy and Nearby Galaxies, 93

\bibitem[{{Chandrasekhar}(1960)}]{chandrasekhar} {Chandrasekhar}, S. 1960,
  {Radiative transfer}, ed. S.~{Chandrasekhar}

\bibitem[{{Drimmel} \& {Spergel}(2001)}]{drimmel01} {Drimmel}, R. \& {Spergel},
  D.~N. 2001, \apj, 556, 181

\bibitem[{{Fish} {et al.}(2003)}]{fish03} {Fish}, V.~L., {Reid}, M.~J., {Argon},
  A.~L., \& {Menten}, K.~M.\ 2003, \apj, 596, 328

\bibitem[{{Gaensler} {et~al.}(2001){Gaensler}, {Dickey}, {McClure-Griffiths},
    {Green}, {Wieringa}, \& {Haynes}}]{gaensler01} {Gaensler}, B.~M., {Dickey},
  J.~M., {McClure-Griffiths}, N.~M., {et~al.} 2001, \apj, 549, 959

\bibitem[{{G{\'o}rski} {et~al.}(2005){G{\'o}rski}, {Hivon}, {Banday}, {Wandelt},
    {Hansen}, {Reinecke}, \& {Bartelmann}}]{gorski05} {G{\'o}rski}, K.~M.,
  {Hivon}, E., {Banday}, A.~J., {et~al.} 2005, \apj, 622, 759

\bibitem[{{Han}(2009)}]{han09} {Han}, J. 2009, in IAU Symposium, Vol. 259, IAU
  Symposium, 455--466

\bibitem[{{Han}(2001)}]{han01} {Han}, J.~L. 2001, \apss, 278, 181

\bibitem[{{Han}(2008)}]{han08} {Han}, J.~L. 2008, in American Institute of
  Physics Conference Series, Vol.  968, Astrophysics of Compact Objects,
  ed. Y.-F. {Yuan}, X.-D. {Li}, \& D.~{Lai}, 165--172

\bibitem[{{Han} {et~al.}(1997){Han}, {Manchester}, {Berkhuijsen}, \&
    {Beck}}]{han97} {Han}, J.~L., {Manchester}, R.~N., {Berkhuijsen}, E.~M., \&
  {Beck}, R. 1997, \aap, 322, 98

\bibitem[{{Han} {et~al.}(2006){Han}, {Manchester}, {Lyne}, {Qiao}, \& {van
      Straten}}]{han06} {Han}, J.~L., {Manchester}, R.~N., {Lyne}, A.~G.,
  {Qiao}, G.~J., \& {van Straten}, W. 2006, \apj, 642, 868

\bibitem[{{Han} {et~al.}(1999){Han}, {Manchester}, \& {Qiao}}]{han99} {Han},
  J.~L., {Manchester}, R.~N., \& {Qiao}, G.~J. 1999, \mnras, 306, 371

\bibitem[{{Han} \& {Qiao}(1994)}]{han94} {Han}, J.~L. \& {Qiao}, G.~J. 1994,
  \aap, 288, 759

\bibitem[{{Han} \& {Wielebinski}(2002)}]{han02} {Han}, J.~L. \& {Wielebinski},
  R. 2002, Chinese Journal of Astronomy and Astrophysics, 2, 293

\bibitem[Han et al.(2004)]{han04} Han, J.~L., Ferriere, K., \& Manchester,
  R.~N.\ 2004, \apj, 610, 820

\bibitem[{{Han} \& {Zhang}(2007)}]{han07} {Han}, J.~L., \& {Zhang}, J.~S.\ 2007,
  \aap, 464, 609

\bibitem[{{Han}(2008)}]{han08b} Han, J.~L.\ 2008, Nuclear Physics B Proceedings
  Supplements, 175, 62

\bibitem[Harari et al.(1999)]{harari99} Harari, D., Mollerach, S., \& Roulet,
  E.\ 1999, Journal of High Energy Physics, 8, 22

\bibitem[Haverkorn et al.(2008)]{haverkorn08} Haverkorn, M., Brown, J.~C.,
  Gaensler, B.~M., \& McClure-Griffiths, N.~M.\ 2008, \apj, 680, 362

\bibitem[{{Haverkorn} {et~al.}(2008){Haverkorn}, {Gaensler}, \&
    {Brown}}]{haverkorn08b} {Haverkorn}, M., {Gaensler}, B.~M., \& {Brown},
  J.-A.~C. 2008, in Mapping the Galaxy and Nearby Galaxies, ed. K.~{Wada} \&
  F.~{Combes}, 329

\bibitem[{{Heiles}(1996)}]{heiles96} {Heiles}, C. 1996, \apj, 462, 316

\bibitem[{{Hinshaw} {et~al.}(2009){Hinshaw}, {Weiland}, {Hill}, {Odegard},
    {Larson}, {Bennett}, {Dunkley}, {Gold}, {Greason}, {Jarosik}, {Komatsu},
    {Nolta}, {Page}, {Spergel}, {Wollack}, {Halpern}, {Kogut}, {Limon}, {Meyer},
    {Tucker}, \& {Wright}}]{hinshaw09} {Hinshaw}, G., {Weiland}, J.~L., {Hill},
  R.~S., {et~al.} 2009, \apjs, 180, 225

\bibitem[Hou et al.(2009)]{hou09} Hou, L.~G., Han, J.~L., \& Shi, W.~B.\ 2009,
  \aap, 499, 473


\bibitem[{{Indrani} \& {Deshpande}(1999)}]{indrani99} {Indrani}, C. \&
  {Deshpande}, A.~A. 1999, New Astronomy, 4, 33

\bibitem[Jansson et al.(2008)]{jansson08} Jansson, R., Farrar, G.~R., Waelkens,
  A.~H., \& et al.\ 2008, International Cosmic Ray Conference, 2, 223

\bibitem[{{Jansson} {et~al.}(2009){Jansson}, {Farrar}, {Waelkens}, \&
    {En{\ss}lin}}]{jansson09} {Jansson}, R., {Farrar}, G.~R., {Waelkens}, A.~H.,
  \& {En{\ss}lin}, T.~A. 2009, Journal of Cosmology and Astro-Particle Physics,
  7, 21

\bibitem[{{Kutschera} \& {Jalocha}(2004)}]{kutschera04} {Kutschera}, M. \&
  {Jalocha}, J. 2004, Acta Physica Polonica B, 35, 2493

\bibitem[{{La Rosa} {et~al.}(2006){La Rosa}, {Shore}, {Joseph}, {Lazio}, \&
    {Kassim}}]{larosa06} {La Rosa}, T.~N., {Shore}, S.~N., {Joseph}, T.,
  {Lazio}, W., \& {Kassim}, N.~E.  2006, Journal of Physics Conference Series,
  54, 10

\bibitem[{{Masip} \& {Mastromatteo}(2008)}]{masip08} {Masip}, M. \&
  {Mastromatteo}, I. 2008, Journal of Cosmology and Astro-Particle Physics, 12,
  3

\bibitem[{{Men} {et~al.}(2008){Men}, {Ferri{\`e}re}, \& {Han}}]{men08} {Men},
  H., {Ferri{\`e}re}, K., \& {Han}, J.~L. 2008, \aap, 486, 819

\bibitem[{Miville-Desch{\^e}nes} {et al.}(2008)]{miville08}
  {Miville-Desch{\^e}nes}, M.-A., {Ysard}, N., {Lavabre}, A., {Ponthieu}, N.,
  {Mac{\'{i}}as-P{\'e}rez}, J.~F., {Aumont}, J., \& {Bernard}, J.~P.\ 2008,
  \aap, 490, 1093


\bibitem[{{Nelson}(1988)}]{nelson88} {Nelson}, A.~H. 1988, \mnras, 233, 115

\bibitem[{{Noutsos} {et~al.}(2008){Noutsos}, {Johnston}, {Kramer}, \&
    {Karastergiou}}]{noutsos08} {Noutsos}, A., {Johnston}, S., {Kramer}, M., \&
  {Karastergiou}, A. 2008, \mnras, 386, 1881

\bibitem[{{Page} {et~al.}(2007){Page}, {Hinshaw}, {Komatsu}, {Nolta}, {Spergel},
    {Bennett}, {Barnes}, {Bean}, {Dor{\'e}}, {Dunkley}, {Halpern}, {Hill},
    {Jarosik}, {Kogut}, {Limon}, {Meyer}, {Odegard}, {Peiris}, {Tucker},
    {Verde}, {Weiland}, {Wollack}, \& {Wright}}]{page07} {Page}, L., {Hinshaw},
  G., {Komatsu}, E., {et~al.} 2007, \apjs, 170, 335

\bibitem[{{Plante} {et~al.}(1995){Plante}, {Lo}, \& {Crutcher}}]{plante95}
  {Plante}, R.~L., {Lo}, K.~Y., \& {Crutcher}, R.~M. 1995, \apjl, 445, L113

\bibitem[{{Poezd} {et~al.}(1993){Poezd}, {Shukurov}, \& {Sokoloff}}]{poezd93}
  {Poezd}, A., {Shukurov}, A., \& {Sokoloff}, D. 1993, \mnras, 264, 285

\bibitem[{{Prouza} \& {{\v S}m{\'{\i}}da}(2003)}]{prouza03} {Prouza}, M. \& {{\v
      S}m{\'{\i}}da}, R. 2003, \aap, 410, 1

\bibitem[{{Rand} \& {Kulkarni}(1989)}]{rand89} {Rand}, R.~J. \& {Kulkarni},
  S.~R. 1989, \apj, 343, 760

\bibitem[{{Reich}(2006)}]{reich06} {Reich}, W. 2006, ArXiv Astrophysics e-prints

\bibitem[Roy et al.(2008)]{roy08} Roy, S., Pramesh Rao, A., \& Subrahmanyan, R.\
  2008, \aap, 478, 435

\bibitem[Rubi{\~n}o-Mart{\'{\i}}n et al.(2010)]{rubino-martin10}
  Rubi{\~n}o-Mart{\'{\i}}n, J.~A., et al.\ 2010, Highlights of Spanish
  Astrophysics V, 127

\bibitem[{{Rybicki} \& {Lightman}(1986)}]{rybicki} {Rybicki}, G.~B. \&
  {Lightman}, A.~P. 1986, {Radiative Processes in Astrophysics},
  ed. G.~B. {Rybicki} \& A.~P. {Lightman}

\bibitem[{{Simard-Normandin} \& {Kronberg}(1980)}]{simard80} {Simard-Normandin},
  M. \& {Kronberg}, P.~P. 1980, \apj, 242, 74

\bibitem[Stanev(1997)]{stanev97} Stanev, T.\ 1997, \apj, 479, 290

\bibitem[{{Strong} {et~al.}(2007){Strong}, {Moskalenko}, \&
    {Ptuskin}}]{strong07} {Strong}, A.~W., {Moskalenko}, I.~V., \& {Ptuskin},
  V.~S. 2007, Annual Review of Nuclear and Particle Science, 57, 285

\bibitem[{{Sun} {et~al.}(2008){Sun}, {Reich}, {Waelkens}, \&
    {En{\ss}lin}}]{sun08} {Sun}, X.~H., {Reich}, W., {Waelkens}, A., \&
  {En{\ss}lin}, T.~A. 2008, \aap, 477, 573

\bibitem[Tauber et al.(2010)]{tauber2010} Tauber, J. et al. 2010, \aap, in press
  (DOI: 10.1051/0004-6361/200912983).

\bibitem[{{Testori} {et~al.}(2008){Testori}, {Reich}, \& {Reich}}]{testori08}
  {Testori}, J.~C., {Reich}, P., \& {Reich}, W. 2008, \aap, 484, 733

\bibitem[{{The Planck Collaboration}(2006)}]{planck06} {The Planck
    Collaboration}. 2006, ArXiv Astrophysics e-prints

\bibitem[Tinyakov \& Tkachev(2002)]{tinyakov02} Tinyakov, P.~G., \& Tkachev,
  I.~I.\ 2002, Astroparticle Physics, 18, 165

\bibitem[{{Vallee}(1991)}]{vallee91} {Vallee}, J.~P. 1991, \apj, 366, 450

\bibitem[Vallee(1995)]{vallee95} Vallee, J.~P.\ 1995, \apj, 454, 119

\bibitem[Vall{\'e}e(2002)]{vallee02} Vall{\'e}e, J.~P.\ 2002, \apj, 566, 261

\bibitem[{{Vall{\'e}e}(2008)}]{vallee08} {Vall{\'e}e}, J.~P. 2008, \apj, 681,
  303

\bibitem[{{Weisberg} {et~al.}(2004){Weisberg}, {Cordes}, {Kuan}, {Devine},
    {Green}, \& {Backer}}]{weisberg04} {Weisberg}, J.~M., {Cordes}, J.~M.,
  {Kuan}, B., {et~al.} 2004, \apjs, 150, 317

\bibitem[{{Wolleben} {et~al.}(2006){Wolleben}, {Landecker}, {Reich}, \&
    {Wielebinski}}]{wolleben06} {Wolleben}, M., {Landecker}, T.~L., {Reich}, W.,
  \& {Wielebinski}, R. 2006, \aap, 448, 411

\end{thebibliography}

\end{document}